\documentstyle[epsfig]{l-aa}
\newcommand{\myfigure} [4] {
        \begin{figure*}
	\setbox100=\hbox{
	\epsfxsize=#1 cm
	\epsfbox{#2.ps}}
	\centerline{\hbox{\box100}}
	\caption{#3}
	\label{#4}
	\end{figure*}
}
\newcommand{\dessin} [4] {
        \begin{figure}
	\setbox100=\hbox{
	\epsfxsize=#1 cm
	\epsfbox{#2.ps}}
	\centerline{\hbox{\box100}}
	\caption{#3}
	\label{#4}
	\end{figure}
}

\begin{document}
\thesaurus{ 11.07.1 --   
 11.08.1 -- 
 11.09.1 --  
 11.09.4 --   
 11.11.1 --  
 12.04.1  }   

\title{The 3D Geometry of Dark Matter Halos}
\author{J.-F. Becquaert and F. Combes}
\institute{ DEMIRM, Observatoire de Paris, 61 Av. de l'Observatoire,
 F--75014, Paris, France}
\date{\today}
\maketitle
\begin{abstract}
The thickness of the neutral hydrogen layer,
coupled with the rotation curve, traces the outer dark matter 
potential. 
We estimate the amplitude of the flaring in spiral galaxies 
from a 3D model of the HI gas.
Warps in particular are explicitly parametrized in the form
of an harmonical density wave.
Applying our method to the galaxy NGC 891, the only model that 
could fit the observations, and in particular
the HI at large height above the plane, includes a strong warp with
a line of node almost coinciding with the line of sight. This high-Z HI is
not observed at the most extreme velocity channels, those corresponding
to high rotational velocities. This is accounted for by the model, since
orbits in the tilted planes are not circular, but elongated, with their
minor axis in the galaxy plane. Their velocity on the major axis (i.e.
at their maximal height above the plane) is then 30\% less than in the 
plane. 
We finally connect the modelled vertical outer gaseous distribution to
the dark matter through hydrodynamical and gravitational equations.
Under the assumption of isotropy of the gaseous velocity dispersion,
we conclude on a very flattened halo geometry for the galaxy NGC 891 
($q \approx 0.2$), while a vertical velocity dispersion smaller that 
the radial one would lead
to a less flattened Dark Matter Halo ($q \approx 0.4-0.5$). Both results
however suggests that dark matter is dissipative or has been strongly 
influenced by the gas dynamics.
\end{abstract}
\section{Introduction}
Knowing the 3D shape and extent of 
the Dark Matter Halos embedding galaxies 
(hereafter DMHs) is of particular importance: 
it can constrain the nature of the dark matter itself,
and in particular its dissipative character. Even in 
the frame of a given dark matter model, such as CDM,
the shape of DMHs can constrain the galaxy formation scenarios,
including gas infall (e.g. Dubinski \& Carlberg 1991,
Dubinski 1994). 
Little is known about the shape of DMHs, and in particular
their flattening and extent. Most of the best evidence
of the existence of DMHs has been gained through HI rotation
curves (see the review by Freeman 1993), and are limited to the
HI disk. The dynamics of satellites (Zaritsky \& White 1994)
can extend further but with strong biases. Techniques based on
gravitational lensing, virial masses or X-rays can reach 
constraints at much larger scales. In a recent
paper, David et al (1995) observed a decrease of
the mass-to-light ratio $[M_{tot}/M_{lum}]$ between
galaxies and clusters, suggesting that the dark
matter in the universe is essentially confined around
individual galaxies. 

Since the best tool to test DMHs, HI rotation
curves, only provide azimutal constraints,
we have to find a vertical signature of these 
halos on the plane of spiral galaxies (for which we
have the most accurate rotation curves).
 The knowledge of both the HI velocity dispersion
and plane thickness is a mean to test the
shape of DMHs. Since it has been measured in 
face-on galaxies a constant HI velocity dispersion with radius
(e.g. Dickey et al 1990), the idea is to use
the flaring of the neutral hydrogen (i.e. 
radial increase of the 
thickness of the HI layer towards external parts), as a 
vertical tracer of the dark matter potential.
The HI layer in an isolated exponential stellar
disk is expected to flare exponentially with radius;
 but the amplitude of the flaring is expected to
be modified with the presence of surrounding dark mass, 
depending on its geometry. 
The first serious attempt using the flaring of the gas layers 
has been done by R. Olling (1995)
who concluded on a very pronounced flattening of the halo of 
NGC 4244 of $(\frac{c}{a}=) q = 0.2_{-0.1}^{+0.3}$ (i.e. E5-E9), 
the large error being 
mostly due to the uncertainties on the velocity dispersion 
of the gas.
Other techniques exist: the first one is to constrain 
the flattening using 
Polar Rings (i.e. two ``orthogonal'' rotation curves).
Sackett and Sparke (1990) found using this way a constraint on 
NGC4650a 's DMH of E6-E7 
, while more recently the same study placed a E5 
constraint on A0136-0801's DMH 
(Sackett and Pogge, 1995). However, the method using
Polar Rings is intrinsically uncertain, and the 
Polar Ring itself can totally change the DMH geometry,
as shown by Combes \& Arnaboldi (1995). 
A related method consists in
using the dynamics of a precessing dusty disk: 
 Steiman-Cameron et al (1992) inferred this way for the 
SO galaxy NGC4753 a quasi-spherical E1 DMH.
Deriving the DMH geometry by the flaring law has however 
the advantage to be applicable 
to a larger sample of galaxies : every spiral galaxy with an inclination
 $\geq 60^{\circ}$ can be an acceptable candidate (Olling 1995).\\
The large range of values found until now for DMH flattenings
(from E1 to E9) can be interpreted either in the way that 
the local universe is very heterogeneous, 
or, and this is more probable, that methods quoted above 
are dominated by errors and uncertainties.\\
Determining the flaring of the gas in disk galaxies is a 
difficult task, due to projection uncertainties 
(galaxies are never perfectly edge-on)
and due to the presence of warps, that deform the gas plane 
and thus modifies our perception of its vertical distribution.
We remark that in the literature 
concerning the determination of the flaring of the gas in 
spiral galaxies, warps are never taken into account,
so that the validity of the flaring curves obtained remains 
correct only for the 
inner unwarped disk, precisely where the dark matter halo 
is undetectable: 
a statistical analysis of the morphology of 167 rotation
curves of spiral galaxies indeed shows that the dynamics 
inside the optical regions is
not usually controlled by the dark component 
(Corradi \& Capaccioli, 1990, and Freeman, 1993). 
A predictive and consistent theory of warps still does not exist 
(Binney, 1992)
even if we may have new hints about it (Masset \& Tagger, 1996).
We still do not have a general model of warps, 
an equation connecting the shape of the warp to the fundamental
parameters of the galaxy, so that most of the time, 
the warp is ignored in models. The flaring curves
given in the litterature are therefore overestimated 
(nearly all galaxies being more or less warped).
The DMH flattening is consequently generally underestimated
(see figure 10 of Olling's Thesis, 1995 and our 
Fig~\ref{qdm2}): halos are 
certainly even more ``disk-like'' if we model 
and substract a warp.
More generally and as far as we know, 
no attempt has been made to search different ``geometries''
than ellipsoidal. Combes and Arnaboldi (1995) suggested 
that dark matter could be ring-like, in the same
category of geometry than the Pfenniger \& Combes (1994) 
model of dark matter which is a disk.   
A question arises finally: can we use "warps" as a supplement 
vertical signature ?
The idea that one may take the shape of warps as 
another constraint on the DMH geometry
(Sparke \& Casertano 1988) is more and more controversial.
Tubbs \& Sanders (1979) and Sparke \& Casertano (1988) developped the idea
that the shape of warps can help to constrain the DMH geometry. 
But this is extremely model-dependent. A spherical potential can
maintain a warp forever (Tubbs \& Sanders, 1979) while a self-gravitating
warp also solves the problem of differential
precession (Sparke \& Casertano 1988, Pfenniger et al 1994). 
The HI thickness might thus be the more reliable 
vertical signature of the halo on the outer disk.
\section{Modeling the 3D distribution of the HI gas in spiral galaxies}
In this paper we develop a theoretical method to derive constraints on the DMH
flattening from the observed HI thickness and we apply it to the nearby 
galaxy NGC 891, which is almost edge-on, and on which an abundant 
literature exists.
The HI data of this galaxy used here are from Rupen (1991).
NGC 891 is classified as an Sb galaxy, member of the NGC 1023 group,
at a distance of 10.0 Mpc (Rupen 1991, for H$_0 = 75 \ km/s/Mpc$) and of systemic
velocity 535 km/s.  
This galaxy is not obviously warped, and it has a gaseous
large-scale asymmetry (southern extension up to 
$R_{HI,south} = 26.5$ kpc), while in the 
north the gas stops surprisingly much before the optical radius
($R_{HI,north} = 18.375$ kpc while $R_{opt} = 22.3$ kpc).
Our subject of interest is to determine if there is indeed
a warp, then estimate the true HI thickness and amplitude of the flaring.
The entire modeling is geometrical and kinematical, so that 
the question of assuming any time-dependency
of warps or the spiral arms do not raise here. We give ourselves
a 3D HI model and adapt the parameters to the observations, 
we do not simulate an evolution of the system as does the N-Body technique. 
\subsection{Modeling the HI infinitely thin disk from the integrated 
flux along the line of sight} 
The gaseous surface density can be represented by 
the real part of the Fourier development:
\begin{eqnarray*}
\Sigma_g(R,\Theta) & = & A_0(R) + \sum_{i=1}^{\infty} A_m(R) 
e^{i(k_m R - m (\Theta - \Theta_m))} 
\end{eqnarray*}
For quasi edge-on galaxies, because of the projection, the
$\Theta$ information vanishes, and spiral arms, in the best case, can be
rebuilt with spectral cube modeling (section 4), but not found with
the technique used here.
The technique is to consider the zero moment map of a nearly edge-on galaxy 
(i.e. the integrated HI flux). 
The r.m.s. noise in the channel maps of NGC 891, associated with the synthesized
beam of $FWHM = 20'' \times 20''$, is 0.780 mJy/beam. The velocity resolution, 
after Hanning smoothing, is 20.7 km/s.
The integrated HI flux is obtained in making a slice exactly along the major 
axis of the galaxy providing a curve: g(X) (full line in Fig~\ref{a0n} and 
Fig~\ref{a0s})
, where X is the radius along the major axis (corresponding to the X coordinate
of the natural cartesian frame of the galaxy).\\ 
This curve, g(X), is connected to the radial gaseous distribution $\Sigma_g(R)$
through an integral that can be inverted assuming low optical
depth (Abel inversion formula), yielding:
\begin{eqnarray}
\Sigma_g(R) & = & -\frac{1}{\pi} \int_{R}^{\infty}
                     [\frac{\delta g(X)}{\delta X}] \frac{dX}
                     {\sqrt{X^{2} - R^{2}}}
\end{eqnarray}
\dessin{8}{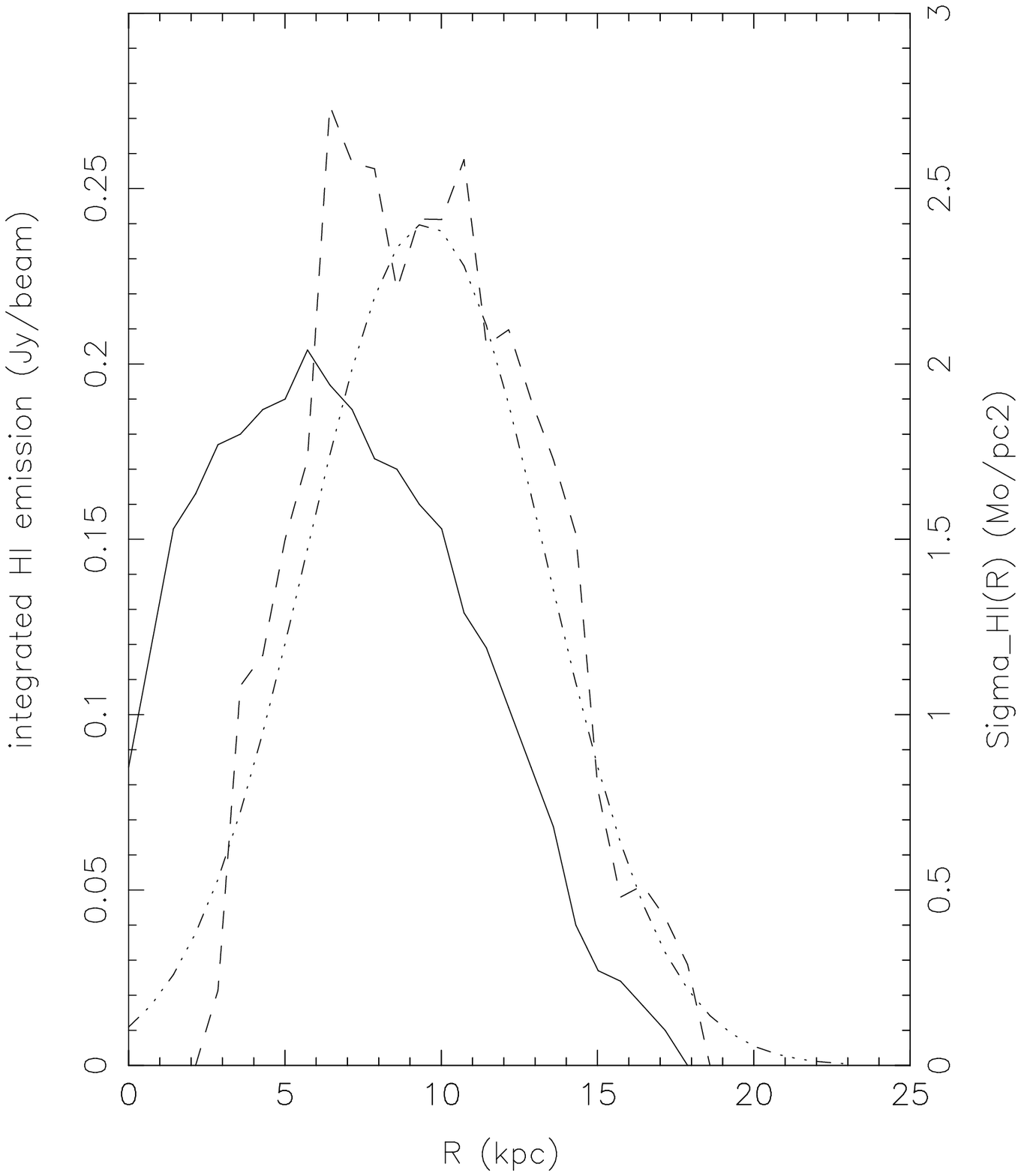}{The full line represents the observational integrated HI flux
along the northern part of NGC 891 from r.m.s corrected zero moment map (in Jy/beam),
 while the dashed line is the surface gaseous density inverted from the integrated
 HI flux assuming low optical depth (in $M_{\odot}/pc^{2}$). The dot-dashed-dot
gaussian curve represents the best fit of the axisymmetric gaseous density ($m=0$),
taking the southern part into account.}
{a0n}
\dessin{8}{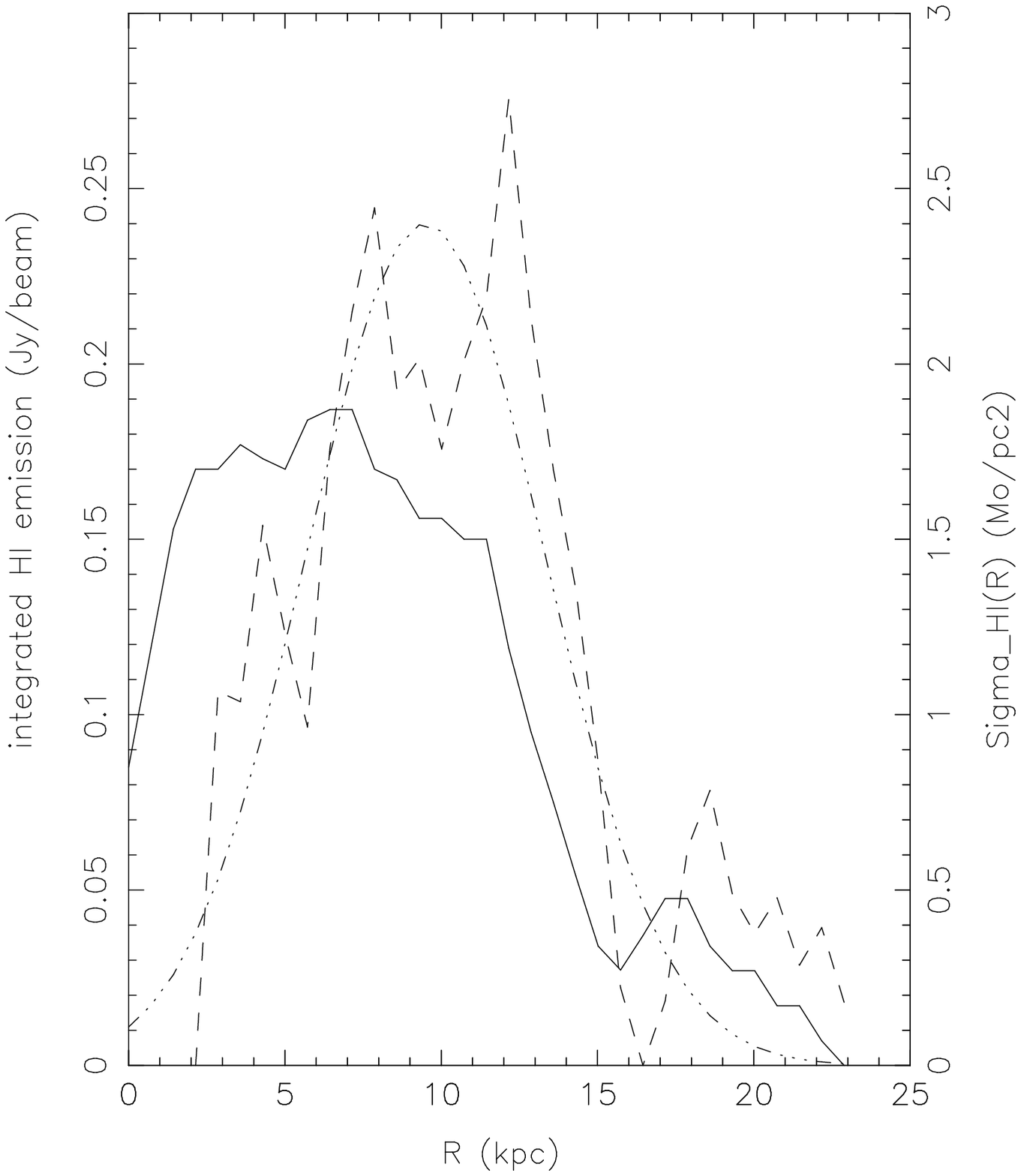}{Same as Fig~\ref{a0n} but for the southern part of NGC 891. We can 
observe a relatively good symmetry with the northern part until 5.5 arcmin 
(fitted with the gaussian curve), confirming the predominance of the axisymmetric 
mode.}
{a0s} 
The results are shown (dashed curve) in Fig~\ref{a0n} and Fig~\ref{a0s}.
We note a good symmetry of the two curves inside $R_{HI,north}$ 
so that at first approximation we may write 
$\Sigma_g(R,\Theta)=\Sigma_g(R) = A_0(R)$.
We fitted the gaseous radial distribution quite well with a gaussian law
of characteristics given below. 
There are two incertainties in this determination. First, the density formula
(1) give negative gaseous densities in the very central parts ($R \le 2$ kpc),
since the derivative of the integrated flux curve is positive, such a 
behaviour of the $g(X)$ curve being due to an HI depletion in the center of the
NGC 891 galaxy, which is quite often seen in the central parts of
spiral galaxies. Second, the integrated flux has an intrinsic error due
the non-infinite resolution, and we could observe that the density inversion 
formula was highly sensitive to local variations of the integrated flux curve.
Assuming that the HI is distributed in a disk of radius $R_{HI,0}$
consists in replacing in the right term of (1) $\infty$ by $R_{HI,0}$.
It was interesting to find that the resulting radial density of the HI gas remains
practically the same with or without this assumption (for $R_{HI,0}=19$ kpc). 
This suggests that the high-Z gas surrounding the galaxy, which is probably 
located at large radii, is no more confined in the thin central HI disk, 
finally suggesting the presence of a warp along the line of sight, which is
also the direction of integration of the HI flux.  
The gaussian law fitting axisymmetrically the northern and 
southern part of NGC 891 is:
\begin{eqnarray}
 A_0(R)     & = & \frac{2.354 \rho_{0,HI}}{\varrho_g \sqrt{2 \pi}}
                  \exp(-\frac{(R - R_g)^{2}}{2 \varrho_g^{2}}) 
\end{eqnarray}
with $R_g = 9.5$ kpc, $\varrho_g = 3.8$ kpc and 
$\rho_{0,HI}=9.6 \ M_{\odot}/pc^{2}$.\\
This fit of the axisymmetric radial gaseous distribution is 
quite sufficient for the rotation curve fit (next section), but for spectral 
cube modeling, we kept integrally the dashed curve representing the "real" 
$\Sigma_g(R)$, through stored data.
We finally calculate the relation between the total HI mass and a gaussian
surface density:
\begin{eqnarray*}
 M_{HI}     & \approx & 2 \pi \rho_{0,HI} \varrho_g R_g 
                    (\sqrt{\frac{\pi}{2}} + 2 \sqrt{\frac{2}{\pi}}) 
\end{eqnarray*}
This gives for $M_{HI} = 4.2 \ 10^{9} \ M_{\odot}/pc^{2}$ (Rupen 1991 for 
$h=0.75$) the value $\rho_{0,HI}=8.0 \ M_{\odot}/pc^{2}$, in 
good agreement with our value of $\rho_{0,HI}=9.6 \ M_{\odot}/pc^{2}$.
\subsection{Modeling the warping of the disk} 
The warp is modelled as a vertical density 
wave of equation (m=1):
\begin{eqnarray}
Z_{warp}(R,\Theta) & = & \Gamma(R) \cos(\Theta - \Theta_w)
\end{eqnarray}
The amplitude of this vertical wave, $\Gamma(R)$, is  
linearly parametrized  but we also
tested a law in $R^{2}$ (paraboloid):
\begin{eqnarray*} 
\Gamma(R \le R_w) & = & 0\\
\Gamma(R \ge R_w) & = & \alpha_{lin} (R - R_w)\\
\Gamma(R \ge R_w) & = & \alpha_{par} (R^2 - R_w^{2})
\end{eqnarray*} 
This parametrization corresponds to a flat
disk's plane until a radius $R_w$, where the warp starts, and
after follows the shape of the $\Gamma$ function.
We primarily included a global wave number for the warp wave 
(a $k_w R$ term in the cosine) but we finally neglected it
for two reasons: firstly because
the line of nodes of the warps are generally observed to be almost 
straight until large radii (Briggs, 1990) yielding that 
$k_w=0$, and secondly because a 
twist of the line of nodes ($k_w \not = 0$), starting after some 
radius, is anyway hard to detect with spectral cube modeling (section 4). 
Simulating modes of warps embedded in dark matter halos,  
Sparke \& Casertano (1988) frequently found linear solutions for the shape 
of the $\Gamma$ functions. 
Briggs (1990) determined the tip angles (i.e. $\Gamma(R)$) 
of warps of a sample of 12 galaxies. Nearly all the features derived 
follow globally a linear or paraboloidal shape. 
Warps are generally observed to raise up at 3-5 exponential scale
lengths of the light distribution, which corresponds more or less to
the location of the optical radius. 
We show in Fig~\ref{zirc90} a 3D model of NGC 891 which revealed to possess 
a warp of high amplitude, beginning at 2.5 exponential scale length (12 kpc).
\subsection{Modeling the HI vertical distribution} 
\dessin{8}{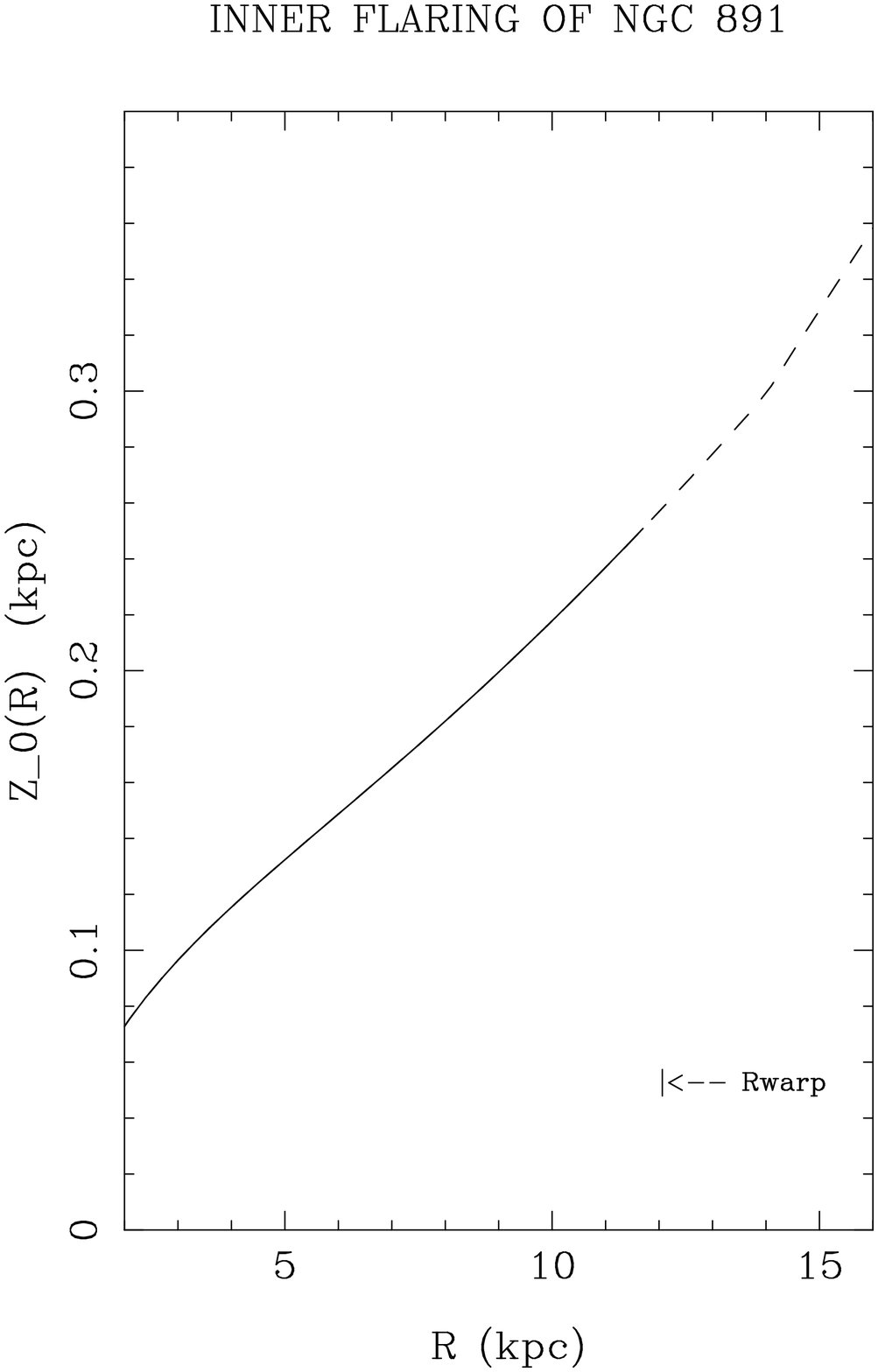}{NGC 891 flaring for a gaussian vertical distribution
from the equation of hydrostatic equilibrium, using the rotation curve
fit section results. This curve is invariant for Z between 0.1 and 0.5 kpc.
Rwarp is determined in the spectral cube modeling section.}
{flaring} 
It is expected from hydrostatic equilibrium 
that the HI layer is vertically
distributed under a gaussian law only if the gas layer is much thinner
than the stellar scale height. For the case of NGC 891, the exponential
stellar disk has a vertical scale height $h_Z \approx 1.0$ kpc, and we
determine in a further section, using our 3D spectral cube modeling, 
that the height of the inner gaseous layer $H_0$ is about 0.1 kpc 
so that we can assume a gaussian vertical distribution 
for the HI gas:
\begin{eqnarray}
\rho_{gas} &\propto & \exp (-\frac{Z^{2}}{2 Z_0(R)^{2}})
\end{eqnarray} 
If we write like Olling (1995) the Hydrostatic Equilibrium of the HI gas,
under the assumption of Z-isothermality and axisymmetry, we obtain:
\begin{eqnarray}
\sigma_{Z}^{2} \frac{d \ln \rho_{HI}}{d Z} & = &
 -\frac{\partial \Phi_{galaxy}}{\partial Z}
\end{eqnarray}
so that inserting (4) in (5) yields: 
\begin{eqnarray}
Z_0(R) & = & \sigma_Z \sqrt{ \frac{Z}{
             \frac{\partial \Phi_{galaxy}}{\partial Z}(R,Z)}}
\end{eqnarray}
Equation (6) is correct for small Z, and R smaller
than $R_w$, since after this radius, the gas rises up in the warp. 
Equation (4) is then modified. Its form is also
dependent on the dark matter model adopted: a disk-like dark 
matter distribution, following the HI layer, settles the gas in a potential well 
and the resolution of equation of the hydrostatic equilibrium will yield a solution
in sech-2, while a spherical dark matter halo will provide a vertical force
proportionnal to Z and thus a gaussian solution for the equation. 
But a sech-2 function and a gaussian have comparable behaviours for Z smaller
than 1 kpc, so that in practice they cannot be separated. We thus adopted a
 gaussian representation:
\begin{eqnarray}
\rho_{gas} &\propto & \exp (-\frac{(Z-Z_{warp}(R,\Theta))^{2}}{2 Z_0(R)^{2}})
\end{eqnarray} 
We derive the galaxy potential of (6) from the rotation curve fit of 
next section.  
We determine using Fig~\ref{flaring}:
\begin{eqnarray*} 
Z_0(R \le 12 \ kpc) & = & 0.08 + 0.014 R \ [kpc]\\
\end{eqnarray*}
We determine further (spectral cube modeling
section) that this galaxy is warped relatively early in 
radius (beginning at about 12 kpc from the center 
of the galaxy, similarly to the Milky Way) so that 
an outer flaring cannot be determined only
with a rotation curve fit.
We observe that the inner flaring obtained above  
is almost linear as expected: many authors,
(Pfenniger et al, 1994 or Olling, 1995) predicted such a behaviour.
This leads us to consider for the spectral cube modeling section 
a warped gaussian HI vertical distribution with a flaring 
of the form:  
\begin{eqnarray} 
Z_0(R \le R_f) & = & H_0 + \zeta_1 R \nonumber \\
Z_0(R \ge R_f) & = & H_0 + \zeta_1 R_f + \zeta_2 (R-R_f)
\end{eqnarray}
where $\zeta_1$ is the inner flaring coefficient 
and $\zeta_2$ the outer flaring coefficient. 
The $\zeta$ coefficients are the slope of the flaring, they
are dimensionless. The $\zeta_2$
coefficient is naturally the most sensitive to the DMH geometry.
This parameter can not be satisfactorily approached 
without modeling in 3 dimensions the warped neutral hydrogen layer.  
\section{Fit of the rotation curve}
\dessin{8.5}{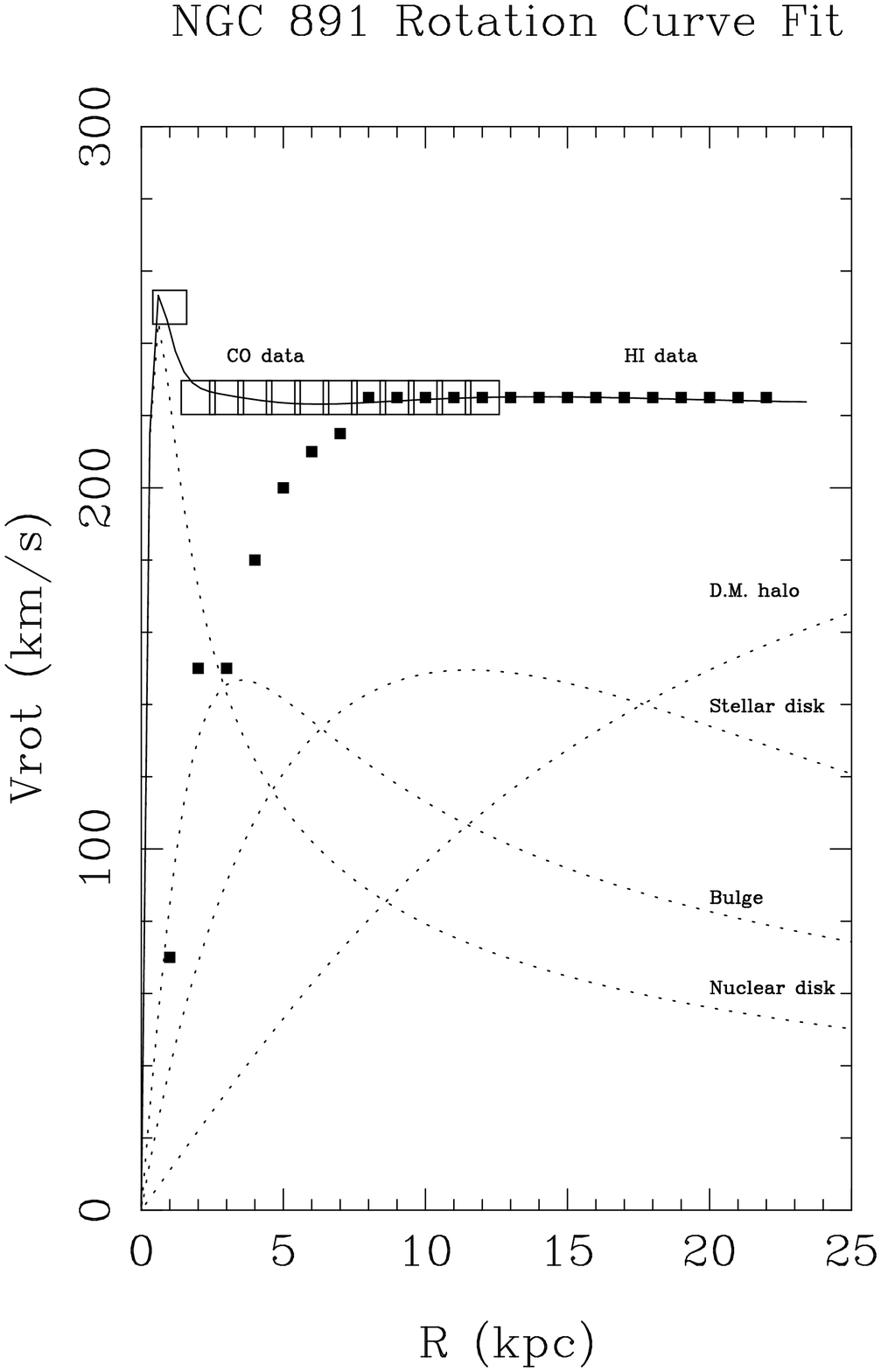}{Multicomponent fit of the rotation curve. Open squares are CO data
from Garcia-Burillo et al (1992) while little full squares are HI data from 
Sancisi \& Allen (1979). For
this fit the dark component has a flattening of $q=0.5$ and a core radius of 14 kpc.
The HI contribution, negligible, is not drawn.}
{vfit1} 
\dessin{8.5}{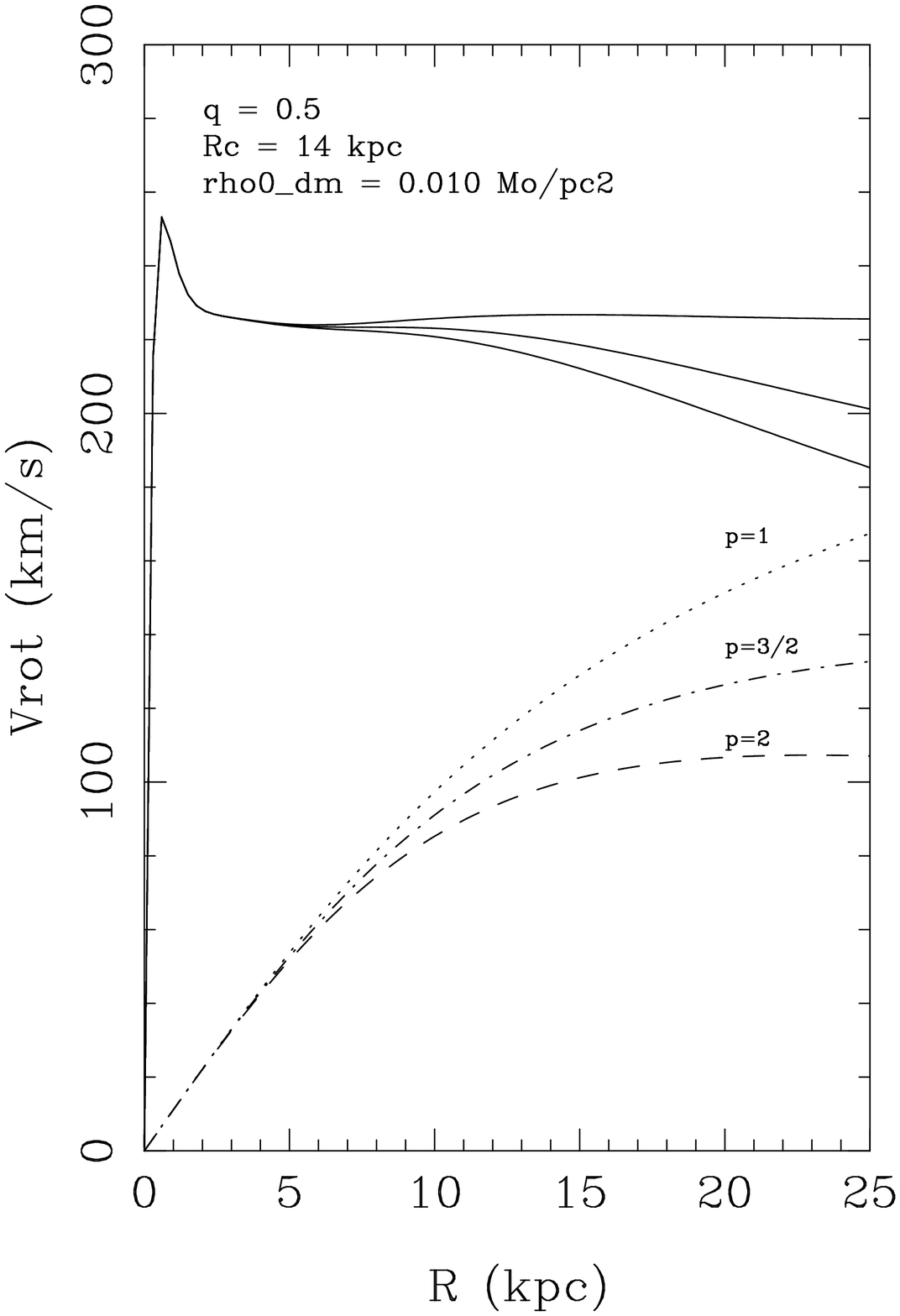}{The impact of modifying the dark matter density
model on the rotation curve fit. Values of the parameters of the 
dark matter halo component are kept fixed, while only the power $p$ changes.}
{power} 
\dessin{8.5}{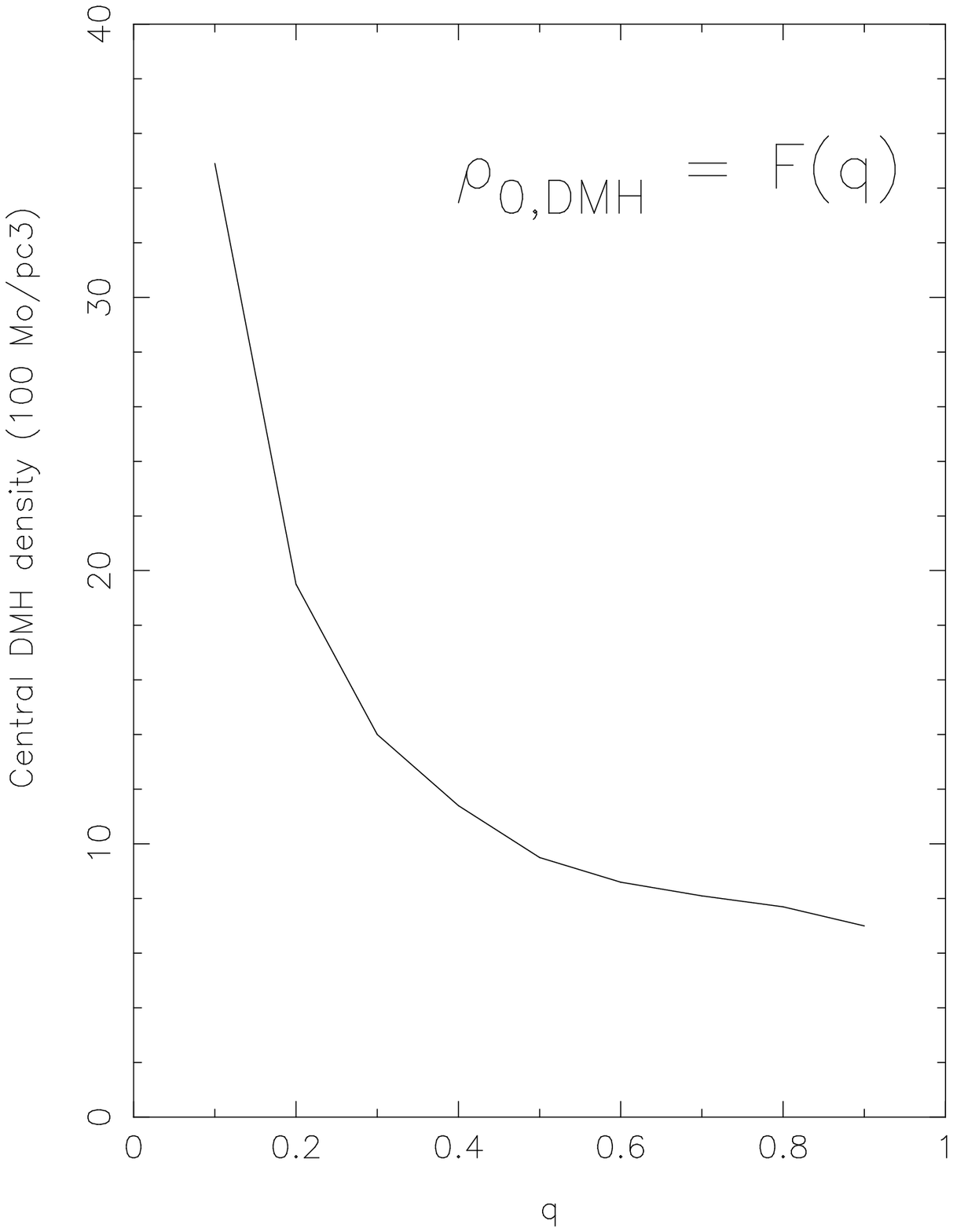}{The central density of the DMH as a function of its flattening, as
found from the rotation curve fit.}
{rofq} 
NGC 891 having a quasi edge-on orientation, and because of HI deficiency
in its center, the HI rotation curve is innacurate inside
8 kpc. The HI points are from Sancisi \& Allen (1979), the relative error being of
about 5 \% after 8 kpc, in the flat region. A multi-component fit, although the
method gives non-unique solutions, requires at least sufficiently good data, so that
we considered for radii smaller than 8 kpc the CO curve provided by 
Garcia-Burillo et al
(1992). NGC891 is expected to have a fast rotating nuclear disk (quite similarly
 to the Milky Way, see Dame et al, 1987), 
explaining why the CO rotation curve steepens to 250 km/s in 500 pc
to fall at 225 km/s and stay constant up to 5 \% at 225 km/s.  
We consider hereafter five components: a Toomre-Kuzmin nuclear disk, a Plummer bulge, 
a thick stellar disk, a gaseous 
contribution calculated for the previously derived radial HI surface density, 
and an ellipsoidal dark matter halo.
The nuclear Toomre disk has a mass $M_{nuclear}$ and a scale-length $s_{nuclear}$.
The Bulge spherical potential is classically
\begin{eqnarray*}
\Phi_B(R,Z) & = &\frac{- G M_B}{\sqrt{R^{2}+ Z^{2} + s_B^{2}}}
\end{eqnarray*} 
and its rotational contribution is straightforward.   
The stellar disk is a thick exponential disk of density (van der Kruit, 1981):
\begin{eqnarray*} 
\rho_{\star}(R \le R_{opt},Z) & = & \rho_{0,\star} \exp (-\frac{R}{h_R})\exp (-\frac{Z}{h_Z})\\
\rho_{\star}(R \ge R_{opt},Z) & = & 0.0
\end{eqnarray*}
The value of $h_Z$ and $h_R$ are kept fixed since several previous studies
have settled them with confidence (see Table 1).
The disk potential is then (Sackett \& Sparke 1990):
\begin{eqnarray*}
\Phi_{\star}(R,Z) & = &
 - G M_{\star}  \int_{0}^{\infty} \frac{J_0(y R) [ y Z_d
 e^{-\mid Z \mid / Z_d} - e^{- y \mid Z \mid} ]}{ 
(1 + h_R^{2} y^{2})^{3/2} (y^{2} Z_d^{2} - 1)} dy
\end{eqnarray*}
We derive the contribution of the stellar disk to the rotation curve:
\begin{eqnarray*}
v_{\star}^{2} & = &
 - G M_{\star} R \int_{0}^{\infty} \frac{y J_1(y R)}{ 
(1 + h_R^{2} y^{2})^{3/2} (y Z_d + 1)} dy
\end{eqnarray*}
We simulated also the equations
of Casertano (1983) who includes explicitly the truncation of the stellar disk.
But considering that the HI data stops before the optical radius in the north, 
and that some HI in the south is probably missing along the line of sight, 
truncation effects on the rotation curve are not expected to be visible, so 
that we adopted the formula of de Zeeuw \& Pfenniger (refound by Sackett \& Sparke, 1990).\\
We finally consider a dark matter component of isothermal, pseudo-ellipsoidal density:
\begin{eqnarray*}
\rho_{DMH} & = & \frac{\rho_{0,DMH}}{(1+ \frac{R^{2}}{R_c^{2}} + 
              \frac{Z^{2}}{R_c^{2} q^{2}})} 
\end{eqnarray*}
This distribution generates the potential:
\begin{eqnarray*}
\Upsilon_{DMH}(R,Z) & = & 2 \pi G \rho_{0,DMH}  q R_c^{2} \\
                   &   & \int_{0}^{\frac{1}{q}}\frac{\ln [1 + \frac{x^{2}}{R_c^{2}} 
(\frac{R^{2}}{\epsilon^{2}x^{2} + 1} + Z^{2})]}{\epsilon^{2}x^{2} + 1}dx
\end{eqnarray*}
where $\epsilon = \sqrt{1-q^{2}}$.
The halo contribution to the rotation curve is therefore:
\begin{eqnarray*}
v_{DMH}^{2} & = & 4 \pi G \rho_{0,DMH} q R^{2} R_c^{2} \\ 
             &   & \int_{0}^{ \frac{1}{q}} \frac{x^{2} dx}
                  {R_c^{2} (\epsilon^{2} x^{2} + 1)^{2} + 
                  (\epsilon^{2} x^{2} + 1) x^{2} R^{2}} 
\end{eqnarray*}
which gives an asymptotic value, at large radii:
\begin{eqnarray*}
v_{DMH,\infty}^{2} & = & 4 \pi G \rho_{0,DMH} R_c^{2} \frac{ q arccosq}{\sqrt{1-q^2}}
\end{eqnarray*}
An interesting particular case of dark matter density model is:
\begin{eqnarray*}
\rho_{DMH} & = & \frac{\rho_{0,DMH}}{(1+ \frac{R^{2}}{R_c^{2}} + 
                 \frac{Z^{2}}{R_c^{2} q^{2}})^{p}} \ (p \ge 0)
\end{eqnarray*}
The corresponding formula for the potential and rotation curve contributions
are given in appendix C. The introduction of the power parameter p (we 
retrieve the normal dark matter density model for $p=1$) is
justified by the work of Lake \& Feinswog (1989), who considered the possibility
that $\rho_{DMH}(R,Z)$ is not necessarily in $R^{-2}$ as it is currently
assumed, the flatness of the rotation curves being also achievable for
profiles in $R^{-3}$ and even $R^{-4}$ (corresponding respectively
to $p=\frac{3}{2}$ and $p=2$). We simulated 
both values of p at the end of this section, 
in order to estimate
the implication on the amount of dark matter required (Fig~\ref {power}).\\ 
Taking the gaussian fit of HI radial distribution,
we derive using formula 2-160 of BT87 the HI contribution:
\begin{eqnarray*}
v_{HI}^{2} & = &  - G \rho_{0,HI} R \int_{0}^{\infty} x J_1(x R)\\ 
           &   &  \lbrack \int_{0}^{\infty} \exp(-
                   \frac{(R - R_g)^{2}}{2 \varrho_g^{2}})  
                   J_0(x y) y dy \rbrack dx
\end{eqnarray*}
In fact, the HI contribution to the rotation 
curve revealed to be negligible: 
in the outer parts, it is of the same order
than the nuclear disk contribution. Inserting the HI contribution
modifies the dark mass required of about 2 \%, less than the
error made in ignoring the stellar truncation.\\
The halo parameter $R_c$ can be somehow constrained
 using the [M/L] ratio of the stellar disk.
Several simulations were made for several halo core radii: in each case
the flatness of the rotation curve could be obtained, but for $R_c$ values
lower than 12 kpc, the disk mass had to be more and more severely 
decreased. The disk mass required for $R_c=14$ kpc is 
$M_{\star} = 7.7 \ 10^{10} M_{\odot}$ 
yielding a mass-to-light ratio in blue color of
$[M/L]_{\star,B} = 7.1 \  M_{\odot}/L_{\odot}$, in excellent 
agreement with the value of 7.0 found by van der Kruit (1981).
For $R_c=8$ kpc, keeping the flatness of the curve requires 
the disk mass to be $M_{\star} = 5.4 \ 10^{10} M_{\odot}$
providing a mass-to-light ratio of 4.9 which is far too low (Bottema et al
 1991).
So that we settle the halo core radius at about $14 \pm 2$ kpc.\\ 
The mass of dark matter enclosed in an oblate ellipsoid of half major axis d
and flattening q is given by:
\begin{eqnarray*}
 M_H(<d)     & = &  4 \pi \rho_{0,DMH} R_c^{2} q
                   [ d - R_c \arctan(\frac{d}{R_c})]
\end{eqnarray*}
Considering the dark mass enclosed within 30 kpc, in a semi-flattened 
ellipsoid of q=0.5, of core radius $R_c=14$ kpc, we calculate that
$M_H = 17 \ 10^{10} M_{\odot}$.\\
The simulation of the generalized dark matter densities with $p=\frac{3}{2}$
and $p=2$ modifies drastically the results. We show in Fig~\ref {power}
the alteration of the dark matter contribution for the same halo parameters,
but for different values of $p$.
Non-isothermal values of $p$ (i.e. $p \ge 1$) require much greater dark matter
central densities, and modifies the stellar disk mass severely. 
We keep $p=1$ for the rest of the article.
We finally remark that fitting the HI data only (i.e. without considering
the inner CO data), leaves unchanged the dark matter component parameters.
The results of our best fit (Fig~\ref {vfit1}) are quoted in Table 1.
We plot in Fig~\ref {rofq} the relation between $\rho_{0,DMH}$ and $q$.
\begin{table}
\begin{flushleft}
\caption{Results of the rotation curve fit.}
\begin{tabular}{c c c c c}
\hline
quantity   &   &         fitting value\\
\hline
 $s_{nuclear}$  & = &  $0.4 \ kpc$ \\ 
 $M_{nuclear}$  & = &  $1.5 \ 10^{10} M_{\odot}$ \\
 $s_B$  & = &  $2.5 \ kpc$ \\ 
 $M_B$  & = &  $3.2 \ 10^{10} M_{\odot}$ \\
 $h_Z$  & = &  $0.99 \ kpc$ \\
 $h_R$  & = &  $4.9 \ kpc$ \\
 $M_{\star}$  & = &  $7.7 \ 10^{10} M_{\odot}$ \\
 $[M/L]_{\star,B}$  & = & $7.1 \  M_{\odot}/L_{\odot}$\\
 $R_c$  & = &  $14 \ kpc$ \\
 $M_H(q=0.5, R \le 30 \ kpc)$  & = &  $17 \ 10^{10} M_{\odot}$\\
 $\rho_{0,DMH}(q=0.1, R_c=14 \ kpc)$  & = &  $0.0349 \ M_{\odot}/pc^{3}$ \\
 $\rho_{0,DMH}(q=0.2, R_c=14 \ kpc)$  & = &  $0.0195 \ M_{\odot}/pc^{3}$ \\
 $\rho_{0,DMH}(q=0.3, R_c=14 \ kpc)$  & = &  $0.0140 \ M_{\odot}/pc^{3}$ \\
 $\rho_{0,DMH}(q=0.4, R_c=14 \ kpc)$  & = &  $0.0114 \ M_{\odot}/pc^{3}$ \\
 $\rho_{0,DMH}(q=0.5, R_c=14 \ kpc)$  & = &  $0.0095 \ M_{\odot}/pc^{3}$ \\
 $\rho_{0,DMH}(q=0.6, R_c=14 \ kpc)$  & = &  $0.0086 \ M_{\odot}/pc^{3}$ \\
 $\rho_{0,DMH}(q=0.7, R_c=14 \ kpc)$  & = &  $0.0081 \ M_{\odot}/pc^{3}$ \\
 $\rho_{0,DMH}(q=0.8, R_c=14 \ kpc)$  & = &  $0.0077 \ M_{\odot}/pc^{3}$ \\
 $\rho_{0,DMH}(q=0.9, R_c=14 \ kpc)$  & = &  $0.0070 \ M_{\odot}/pc^{3}$ \\
\hline
\end{tabular}
\end{flushleft}
\end{table}
\section{Numerical method}
\subsection{The code}
The construction of synthetic spectral cubes
requires the use of many parameters, among which 
six are considered as free parameters: the warp 
parameters $\alpha_w, R_w$ and the flaring parameters 
$R_f, H_0, \zeta_1, \zeta_2$.\\
A rapid code is thus necessary.
 We separate the density model building from 
the kinematics (velocity sampling). To build the density
model, and to project it, we choose to distribute randomly 
particles in space: this technique is much more efficient
(in terms of CPU time)
than a 3D cube construction, for a given spatial resolution;
this is due to the fact that the 3D cube has a very little
volume filling factor, therefore the number of particules required
are much lower than the number of cube cells, and various 3D rotations
are then much quicker.
\subsubsection{Rendering the HI 3D geometry}
The computational
form of the model density is:
\begin{eqnarray*}
 \rho_{HI}(R,\Theta,Z) & = &  \Sigma_g(R)  \ \
                              \exp (-\frac{(Z-Z_{warp}(R,\Theta))
                                ^{2}}{2 Z_0(R)^{2}})
\end{eqnarray*}
where $\Sigma_g(R)$ has been derived from section 2.1, approximated 
by the equation (2), 
$Z_{warp}(R,\Theta)$ is the warp wave described by 
equation (3) and the
flaring of the gas is included in $Z_0(R)$ through the 
$\zeta$-linear law of equation (8).
We generate N particles $[x(i),y(i),z(i); i \in {1,...,N}]$ from
the density model above, on
which we operate a spatial rotation $\Re_{3D} = \Re (I, j)
 \circ \Re (PA, k)$, where "I" is the inclination of the galaxy
and "P.A." its position angle, thus projecting the set of particles 
on the sky plane with the orientation angles of the real galaxy.
\subsubsection{Spectral sampling and synthetic 
spectral cube}
A velocity is associated to every particle in the cube.
The rotational velocity $V_{rot}(R)$ is interpolated from the
previously fitted rotation curve. 
We project the velocities onto the plane of the sky to get radial velocities:
\begin{eqnarray*}        
V_{los} & = & V_{sys} + V_{rot} \sin I \cos \Theta
\end{eqnarray*}
Having a set of particles and the associated radial
velocities, we now sample this 4 dimensions space $[x,y,z,V]$
according to our original NGC 891 spectral cube.
This sampling is done
according to a central (systemic) velocity and a fixed increment 
$\Delta V = \sqrt{\Delta V_{chan}^{2} + \sigma_{HI}^{2}}$
to get channel maps (this assumes an isotropic dispersion $\sigma_{HI}$).
For our test-galaxy, $\Delta V_{chan}$ is 20.7 km/s.
The procedure of the construction is the following.
For the Nst channel map $V_N =V_{sys} + N \times \Delta V_{chan}$, 
we fill a temporary pixel cube with the spatially rotated particles 
in multiplying this geometric cube by a gaussian in $V_N$, centered 
on our previously calculated $V_{los}$, and of dispersion $\Delta V$. 
We finally sums the flux perpendicularly to the plane of the sky
 to obtain a channel map (the Nst). 
We finally loop over N (in our case from 5 to 25, 15 corresponding
the systemic velocity) to get an entire modelled spectral cube
$[X, Y, V_{los}]$. 
The last operation consists in convolving every pixel of
the spectral cube by gaussians of FWHM corresponding to the 
synthetized beam of the VLA observations ($20'' \times 20''$ for 
NGC 891).
\subsubsection{Application of the code}
The random launching of particles in space takes negligible CPU,
in front of the kinematics part. The dimensions of the cube
is 300x300xN$_{chan}$, yielding a spatial resolution of 1 pixel=
3 arcseconds (=147 parsecs), but the technique allows to build
channels one by one, and
only some of the total number of channels (N$_{chan}$=30) are
computed for the whole set of free parameters. 
In practice, 6 channel maps (whatever the number of 
particles) are sampled and beam-smoothed from one spectral
cube in about 30 seconds on one processor of a CONVEX 3440.\\
Spatial resolution is the dominant parameter to the control of
CPU time. 
The observed HI cube of NGC 891 displays 30 channels, 24 of them having
significant flux (channels 2 to 25), the center being the systemic
velocity channel 15 (see Rupen 1991 for the mosaic figure of the 
30 channels). To restrict at maximum the CPU, we modelled every
two channels. The inclination 
of this galaxy has been estimated several times in the literature, 
leading to a firm lower limit
$I \ge 88.6^{\circ}$ (Rupen, 1991). The inclination is thus 
kept constant at 89.0 
degrees since the uncertainty on its value is too small 
($\pm 0.3$ degree) to assign $I$ to be a free parameter.
In the same way, the position angle was fixed to 23.5 degrees
(Garcia-Burillo et al 1992). 
Van der Kruit (1981), using Velocity-Position diagrams of models,
concluded to a constant velocity dispersion of the gas (assumed
isotropic) of 10.0 km/s. 
\subsubsection{The experiments}
A library of models exploring the free parameters are compared to
the HI observations (in figures ~\ref{wx1}, ~\ref{wx2} and ~\ref{wx3}).
With linear contours centered on the half-intensity of the channels, 
the spectral cube of NGC 891 (whether south or north)
do not show any feature resembling to what Olling 
 nicely calls "butterfly wings", 
which are characteristic of HI flaring and warp. 
To make weak structures at large height above the plane emerge,
contours have to be adequately selected, as in Fig~\ref{wx1},
which presents flaring and warped HI layers.\\ 
Our first idea was that the surrounding high-Z gas 
(the low intensity features surrounding the disk up to several kpc height, 
visible in the channel maps "north NGC 891"
of Fig~\ref{wx1}, Fig~\ref{wx2} and Fig~\ref{wx3}) could be interpreted as 
a warp in the line of sight, and so
we fixed the value $\Theta_w = 90^{\circ}$.
We used to test this configuration the northern part 
of the spectral cube since no HI extension along the major axis 
is required.\\
\myfigure{15}{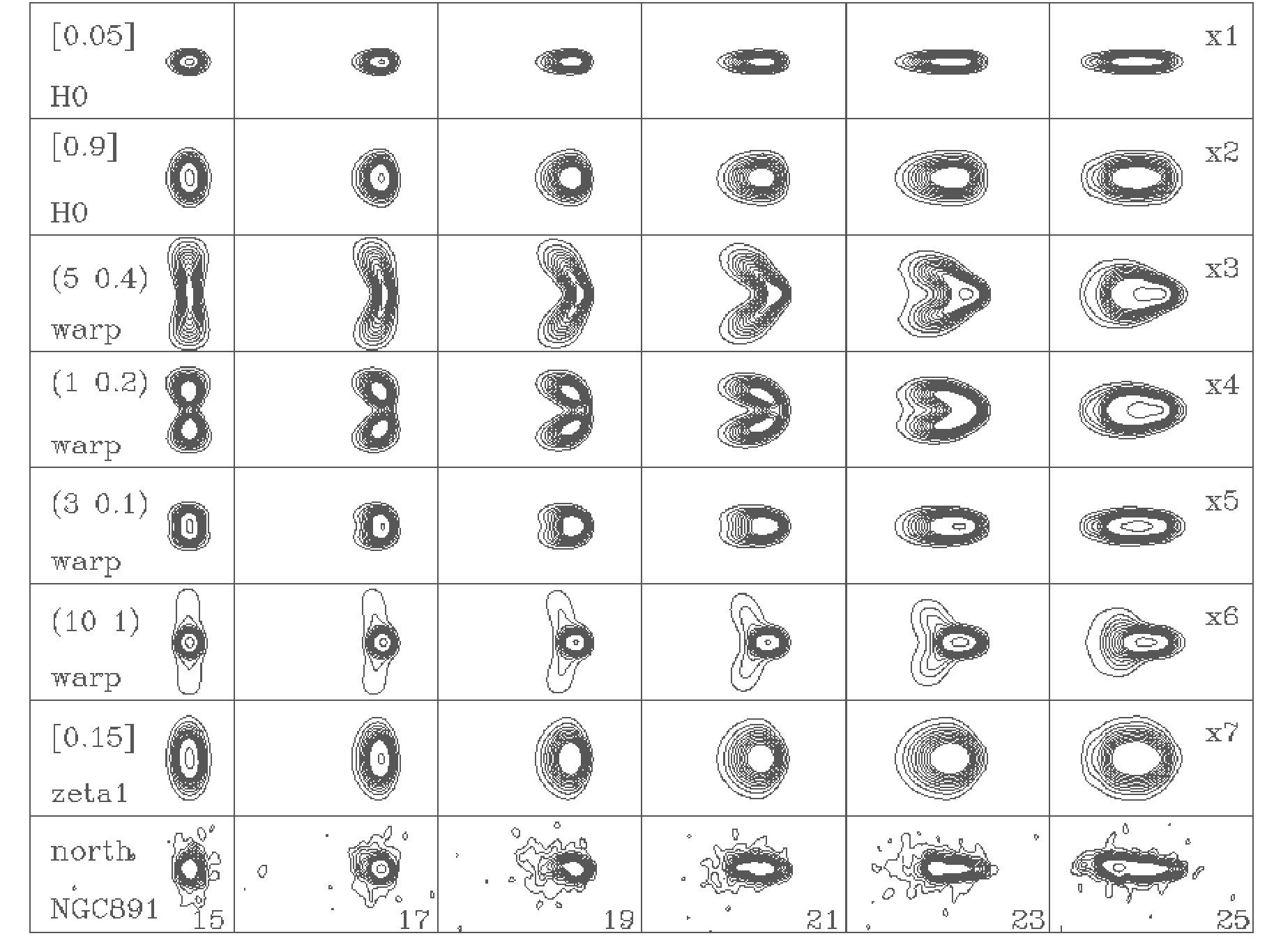}{Library of models: we present here 7 synthetic 
spectral cubes x1 to x7 
(to be compared 
with the original spectral cube "north NGC 891" displayed as the 8th 
cube). 
The same contours are used for every cube:
0.002:0.024:0.002 and 0.036 Jy/beam. The channels presented here are
the systemic channel 15 (535 km/s), and the channels 17 (575 km/s),
19 (620 km/s), 21 (660 km/s), 23 (705 km/s), 25 (745 km/s).}
{wx1}
\myfigure{15}{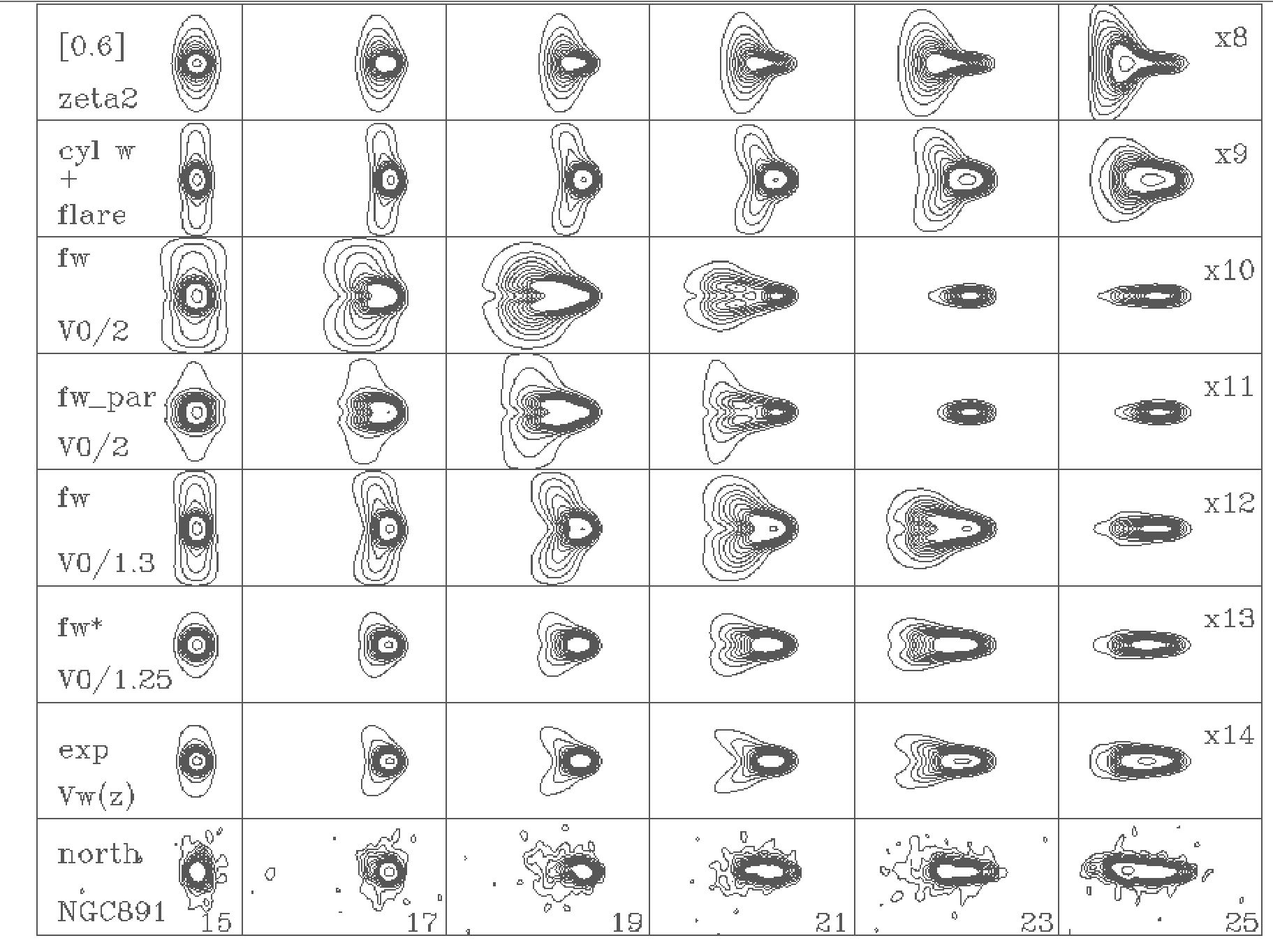}{We present 7 synthetic spectral 
cubes x8 to x14 "converging" towards our best fits x13 and x14
(to be compared with the corresponding channels "north NGC 891"). 
The same contours are used in every cube:
0.002:0.024:0.002 0.036 Jy/beam.}
{wx2}
\myfigure{15}{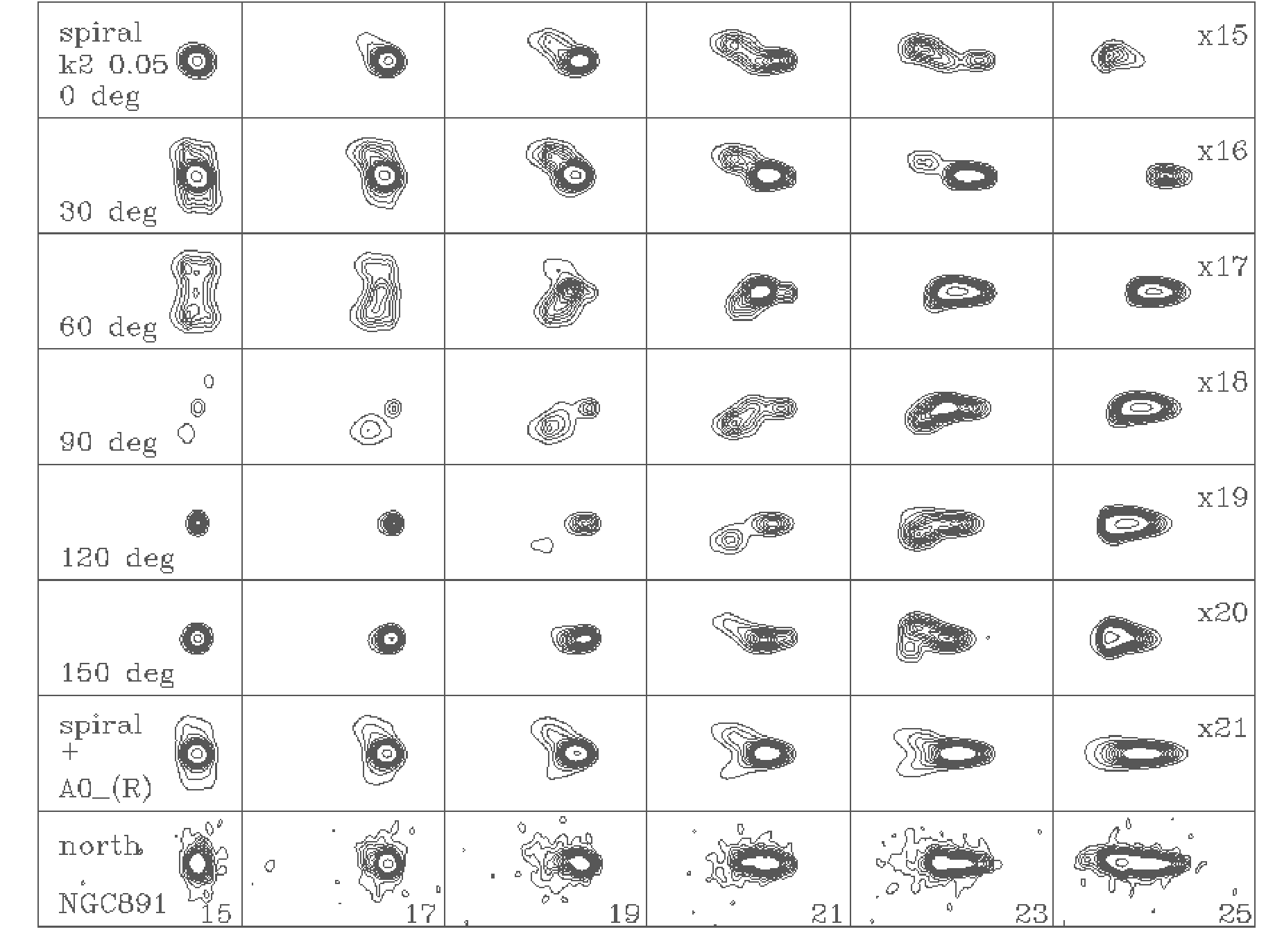}{Spiral cubes: 6 synthetic spectral cubes
representing an $m=2$
spiral mode simulated alone, with a fixed wave number ($k_s=0.05$) 
and a varying line of nodes (from $0^{\circ}$ to $150^{\circ}$). 
The height reached by the spiral has been a bit amplified in settling
$R_{spir}$ at 15 kpc, that is 3 kpc more than $R_w$. We suppose that
$R_{spir}$ is in reality of order $R_w$. 
The seventh cube (x21) is a combination
of a gaussian $A_0(R)$ and an $m=2$ harmonical mode 
with $\Theta_2=30^{\circ}$. 
The same contours are used in every plot:
0.002:0.024:0.002 0.036 Jy/beam.}
{wx3}
The radial distribution of the HI gas
is assumed to be the dashed curve found in section 2.1. 
The gaussian model approximating this dashed curve 
is used later, for the detection of a spiral pattern; from that the
axisymmetric mode is globally determined. 
We see in Fig~\ref{wx1}, Fig~\ref{wx2}, Fig~\ref{wx3} samples of 
our intensive construction of spectral cubes. We vary nearly
all possible parameters, intending to show what transformations on the
2D flux distribution correspond to the variation of each parameter.
The first seven synthetic cubes shown in Fig~\ref{wx1} 
are presented to observe the 2D flux transformations 
that take place for extreme values of 
parameters. We do not search for this first sequence to obtain cubes 
that are close to observations, but rather to build a 
library of models.\\ 
\\
x1: $\alpha_{lin}=0$, $\zeta_1=0$, $\zeta_2=0$, $H_0=0.05$ kpc.\\
\\
x2: $\alpha_{lin}=0$, $\zeta_1=0$, $\zeta_2=0$, $H_0=0.9$ kpc.
This x2 synthetic cube displays a too high HI constant thickness. 
By comparison with the observed cube, the average value is more 
towards low values ($H_0 \approx 0.1-0.5 \ kpc$).\\ 
\\
x3: $\alpha_{lin}(R_w=5 \ kpc)=0.4$, $\zeta_1=0$, $\zeta_2=0$, $H_0=0.4$ kpc.
This synthetic cube displays a warp of too high amplitude, generated too early.
We clearly see that the HI disk flux of the observations is totally
different.\\
\\
x4: $\alpha_{lin}(R_w=1 \ kpc)=0.2$, $\zeta_1=0$, $\zeta_2=0$.
The effect of decreasing the amplitude of the warp, makes its "wings" to 
join. The double "symmetric" holes of the channel 15 are sufficient to rule out
this model.\\
\\
x5: $\alpha_{lin}(R_w=3 \ kpc)=0.08$, $\zeta_1=0$, $\zeta_2=0$.
For very low amplitudes of the warp, we reconstruct the observed HI disk: 
in fact, there
is no more high-Z gas surrounding the disk, as is observed.\\
\\
x6: $\alpha_{lin}(R_w=10 \ kpc)=1.$, $\zeta_1=0$, $\zeta_2=0$.
We show here a warp generated at 10 kpc, further than before.
We observe the emergence of high-Z gas surrounding the central HI disk. 
We also notice that $R_w$ is easy to locate
in the way that the gas after $R_w$ is displaced, inducing a "truncature"
of the outer HI central disk. In particular, the 25th channel is very sensitive
to that modification.
We conclude from comparison with observations that the warp is probably 
generated a little bit further, around $R=12$ kpc.\\
\\
x7: $\alpha_{lin}=0$, $\zeta_1=0.15$, $\zeta_2=0$, $H_0=0.1$.
We observe now the location of an exagerated inner flaring. This 
inner flaring of 0.15 is about ten times the inner flaring expected with the
hydrostatic equilibrium of the gas (section 2.3).\\
\\
x8: $\alpha_{lin}=0$, $\zeta_1=0$, $\zeta_2=0.6$, $R_f=12$, $H_0=0.1$.
Testing the outer flaring coefficient at an abnormal value of 0.6. We observe
that there are clear 2D effects induced.\\
\\
We now try to combine these results to build a sequence of cubes closer
to the observed one.\\
\\
x9: $\alpha_{lin}(R_w=10)=1$, $\zeta_1=0.02$, $\zeta_2=0.02$, $H_0=0.4$.
A flaring warp. We kept the warp parameters of cube x6, in adding a 
flaring of the gas. We notice that the flaring densifies the contours
following the warp along the line of sight, so that even not very well constrained
intrinsically, the flaring can be approached.\\
\\
The main problem when we observe the cube x9 in comparison 
to the observed cube is that the extreme channel 
maps (23 and above all 25) are in every of our models surrounded by 
high-Z gas, while they are not in the observed cube. 
The observed channel map 25 is very thin, 
so that it appears necessary that the particles in the
warp rotate at a different speed than the HI in the disk plane.
This was already noticed by Swaters et al (1997), and interpreted
as non-cylindrical rotation of the HI disk, due to energetic star-formation.
To estimate quantitativally the amount of non-cylindrical rotation,
we adopt a simple kinematical model: we assign a rotational velocity of high-Z
particules lower than in the plane by a free factor (between 1 and 2),
for heights greater than the inner disk
HI thickness, typically 150 parsecs.\\
\\
x10: Simulating high-Z particles as rotating at $V_0/2$, where $V_0$ 
is the "plateau" speed of the rotation curve in the plane (i.e. 225 km/s). 
The parameters for this cube are:\\
$\alpha_{lin}(R_w=9)=1.3$, $\zeta_1=0.02$, $\zeta_2=0.02$, $H_0=0.4$.\\
We observe that the high-Z gas concentrates
around the center channels 15, 17, 19 leaving the extreme channel 23, 25 
without high-Z gas, which is one of the 
characteristics of the observed spectral cube.\\
\\
x11: same parameters but now with a parabolic warp instead of linear, 
of same maximum height as the previous cube x10.
We observe by comparison that a linear warp makes "round" waves 
instead of "square" waves obtained with a $\Gamma$ function in $R^{2}$. 
A linear warp seems closer to the observations. 
We also can appreciate that 
the warp is generated too early, truncating the end of the central
HI disk (visible in particular for the 25th channel).\\ 
\\
x12: same as x10, but with $V_{warp} = V_0/1.3$.
We see that it is possible to make the 25th channel very thin,
as in the observed cube. 
A velocity ratio of 1.3 shows up to be closer to the 
observations than the velocity ratio of 2 of the previous model cube since
it allows the 23rd channel to have high-Z gas.\\ 
\\
x13: A more accurate cube with $V_{warp} = V_0/1.25$. We have here the 
definitive values of the geometric parameters:\\
$\alpha_{lin}(R_w=12)=0.8$, $\zeta_1=0.02$, $\zeta_2=0.02$, $H_0=0.1$\\ 
\\
x14: We refine the kinematical description of the
warp behaviour detected. Instead of considering that the warp cylindrically
rotates, we test now a vertical velocity gradient with an empirical
 "exponential" law. The model becomes:
\begin{eqnarray*}        
V_{los} & = & V_{sys} + V_{rot} \sin I \cos \Theta \exp(-0.1 |Z|)
\end{eqnarray*}
\\
Regarding the hypothesis of the model
(as the decoupling R-Z of the model) and the fact that the radial distribution 
of the gas in section 2.1. is not perfect,
we consider as successful (with the 2 last synthetic spectral cubes x13 and x14) 
our goal to rebuild the observations.\\
\\
We try now to simulate the non-symmetric features observed in NGC 891
cube. We consider the sum of the R-gaussian axisymmetric mode $A_0(R)$ determined 
in section 2.1 and an $m=2$ harmonic term. We generate the spiral after the HI
central depletion of the galaxy, at about 2 kpc, let it develop and make it decrease
in $1/R$ after a radius $R_{spir}$.  The formulation in the model is the following
\begin{eqnarray*}  
\Sigma_g(R \le 2 \ kpc) & = & 0 \\
\Sigma_g(R \le R_{spir}) & = & A_0(R) + A_2 \cos(k_2 R - 2(\Theta-\Theta_2)) \\
\Sigma_g(R > R_{spir}) & = & A_0(R)+
                        \frac{A_2 \cos(k_2 R - 2(\Theta-\Theta_2))}{R-R_{spir}}               
\end{eqnarray*}
where $k_2$ is the global wave
number, $\Theta_2$ the line of nodes of the spiral pattern, and where 
$A_2$ is of small magnitude in comparison to $A_0(R)$. 
Computing alone (i.e. $A_0(R)=0$) the $m=2$
term in our spectral cube code (Fig~\ref{wx3}), we could associate the
external "torsion" of the gas, unexplained with our symmetric models, 
to an $m=2$ warped spiral. We kept the flaring warp parameters
found in the axisymmetric mode construction. The velocity of the
spiral arms is calculated in Appendix B and is implemented in the code. 
Synthetic cubes x15 to x20 show an $m=2$ spiral 
with $k_2=0.05$ and regularly varying line of nodes (from $0^{\circ}$ to 
$180^{\circ}$). It is important to
notice that $R_{spir}$ is found greater than $R_w$ (of about 3 kpc), 
explaining the asymmetric features of the high-Z gas: the spiral is raised by the warp
before its density vanishes, so that a non-negligible quantity of matter 
is asymmetrically placed in the line of sight, at heights corresponding 
to the warp wave.\\
\\
x21: This synthetic cube is the superposition of the axisymmetric
gaussian density found in section 2.1., with an $m=2$ spiral
($k_2=0.05$, $\Theta_2=30^{\circ}$). 
We observe unambiguously that a spiral 
can reproduce the observed asymmetry. It is noteworthy that the
gaussian HI density is only an approximation in front of 
the "real" $\Sigma_{HI}(R)$
stored in buffers: the 25th channel, as an example, is less well fitted
(see x14 for comparison).
\subsubsection{Conclusion on the HI 3D distribution according to our model}
Several hundreds of synthetic cubes under the configuration of a warp
along the line of sight ($\Theta_w=90^{\circ}$) were finally built.
Most parameters induce typical and particular 2D effects:\\
1- a flaring warp along the line of sight makes "butterfly-shapes" that can be
associated to the observed high-Z features of the gas.\\
2- a flaring also thickens regularly the central HI disk, 
a behaviour which is clear for the 23rd channel of NGC 891.\\
Combinations step by step of the parameters 
succeeded to reproduce most of the features observed in the 
original spectral cube of NGC 891. The radius where the warp
begins, for instance, is very sensitive: we are able to set it 
with a good confidence at $12 \pm 0.3$ kpc. The amplitude of the
warp, $\alpha_{lin}$, is about 0.7 to 0.8, depending
on the true pattern of the $m=2$ spiral. The most important result is that 
we found possible to 
constrain the flaring (see synthetic cubes with varying 
$\zeta_1, \zeta_2$ coefficients).
Let us summarize the other deductions:\\
1- The flaring found after modeling in 3D is
of the same order than the value estimated using 
the potential derived from the results of the rotation 
curve fit section (section 2.3).\\ 
2- Modeling with success the gas in 3D requires 
the use of every channel. The central channels were needed in our case
to constrain the amplitude of the warp, and the extreme channels 
23 and 25 were needed to detect the kinematical behaviour of the
warp wave. In particular, the channels 15 and 23 showed to be the 
most sensitive to the flaring coefficients.\\
3- The kinematical behaviour of the warp appeared to be 
dominated by an important velocity gradient.
\dessin{10}{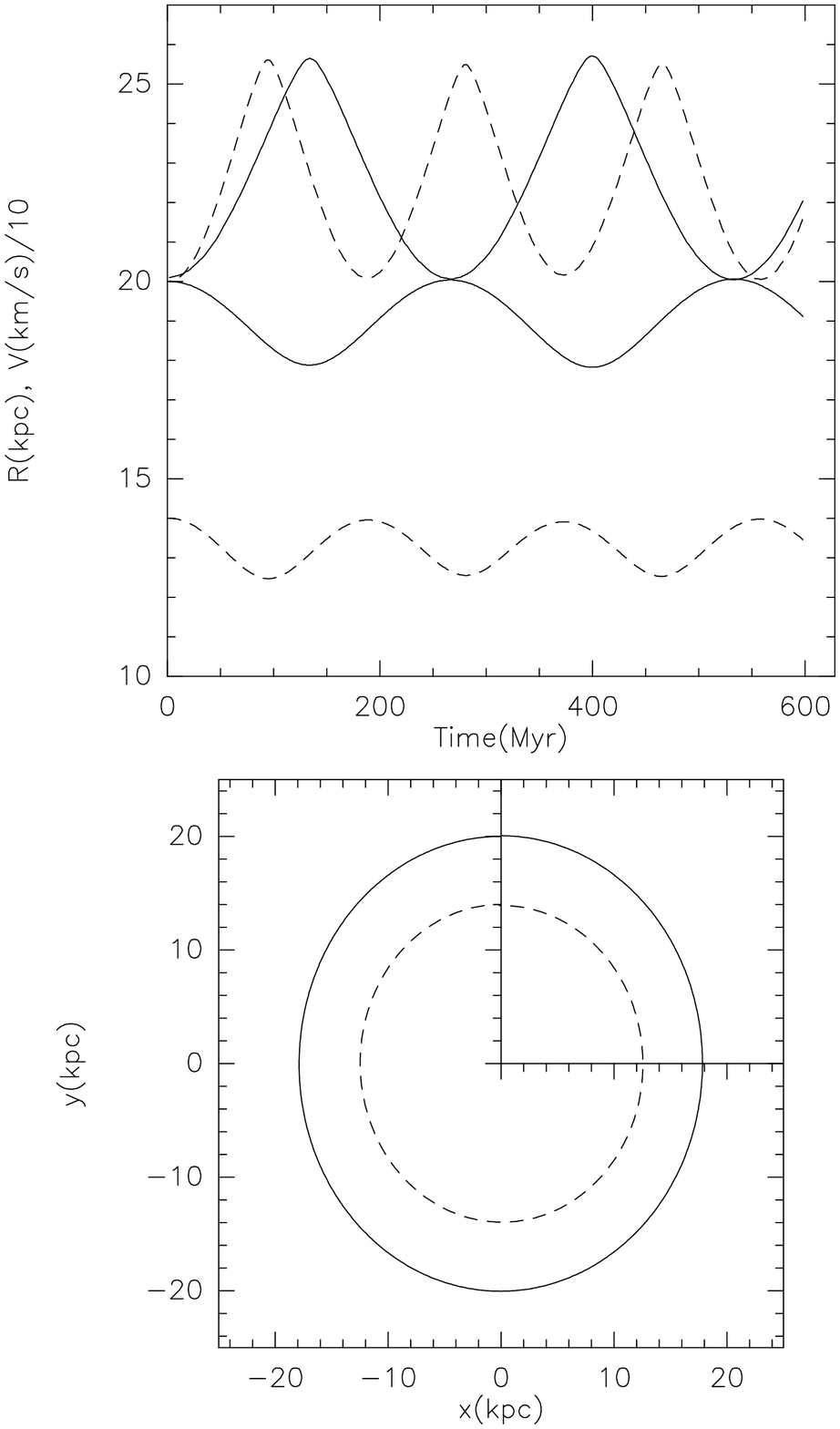}{Computation of the orbits with the shooting method;
{\it bottom}: shape of closed orbits launched at R=14 and 20 kpc, in a plane
tilted by 20$^\circ$ from the equatorial plane, the major axis is
along Ox;
{\it top}: Time evolution of the radius and velocity of the two orbits,
one launched at 20 kpc (full line), the other at 14 kpc (dashed line).
 The largest relative variations are those of the velocity (about 30\%)}
{orbits}
\myfigure{16}{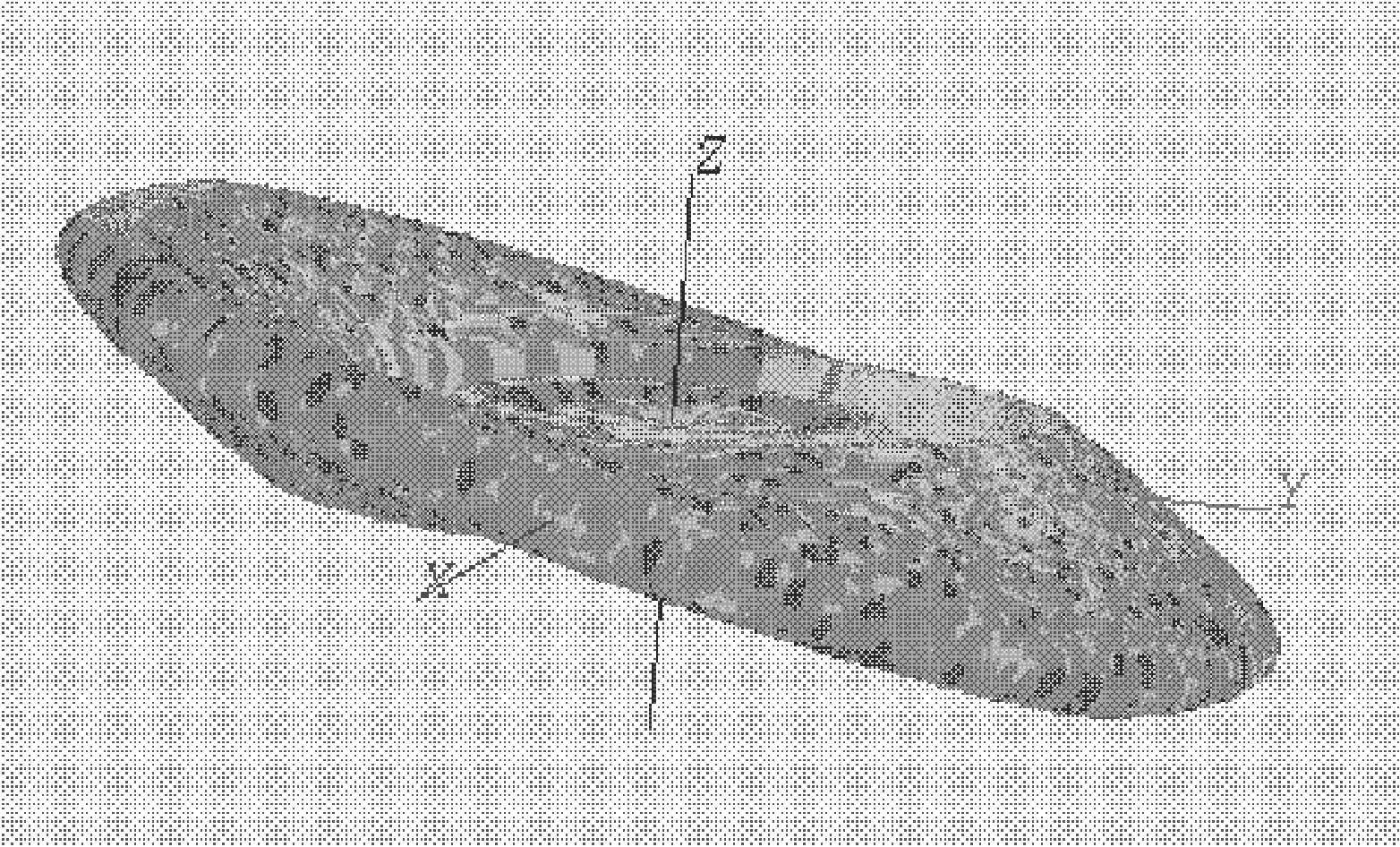}{The 3D HI isodensity contour of NGC 891 
from a synthesis of our results. The radial distribution comes
from section 2.1 while the vertical gaussian distribution has the parameters
found in the spectral cube modeling: $H_0=0.1$ kpc, $\zeta_1=\zeta_2=\zeta=0.02$.
The warp appears clearly along the Y axis accordingly to $\Theta_w=90^{\circ}$.}
{zirc90}
If there is indeed a strong warp, so that the HI gas reaches a height of a few
kpc above the plane at a radius of 20 kpc, then we expect a significant
rotational velocity gradient perpendicular to the plane. This gradient
is of much larger amplitude that the simple gradient expected from
non-cylindrical rotation (appendix A). 
It is due to the fact that the gas orbits in
inclined planes (as in the tilted ring model of warps) while the bulk
of the matter, represented by the stars, is flattened in the plane. The
periodic orbits in the tilted plane are not circular, but elongated,
with their minor axis in the plane of the galaxy. Their tangential 
velocity is then larger on the minor axis (i.e. in the plane) than 
on their major axis (at their apocenter, at their maximum height above
the plane). This produces a z-velocity gradient of the measured rotational
velocity. To compute the order of magnitude of this gradient, we have
taken our model of the NGC 891 potential, including the bulge,
nuclear disk, and mainly the exponential stellar disk.
We also introduced the dark halo, with variable flattening, but this
had only a small influence on the orbits between 10 and 20 kpc.
 To find the closed elongated orbits in the tilted planes,
 we then used the shooting method, as in the computations of the
kinematics of polar rings (e.g. Sackett et al 1994, Combes \& Arnaboldi 1996). 
We discovered that, for all orbits between 10 and 20 kpc average radii,
the average ellipticity of the orbits is $\sim$ 12\%, and the tangential
velocity on the major axis is $\sim$ 30\% less than that on the minor
axis, for a tilt of the plane of 20$^\circ$
(see figure \ref{orbits}). By comparison, the expected
non-cylindrical z-velocity gradient for the same tilt is of the order of
10\% only.
Assuming such a warp for NGC 891, with a line of nodes coinciding
with the line of sight, permits us to fit the observations quite nicely.
 The HI above the plane on the minor axis is taken into account, with
a rotational velocity less than in the plane, which removes high-Z HI
from the extreme velocity channels. The high-Z HI is then confined in 
channels closer to the systemic velocity, as shown in Fig~\ref{wx2}.
 We can note a posteriori that this effect can produce an over-estimation
of the rotational velocity at large radii, since the velocity in the 
plane is higher than it should be for circular orbits. It is then
possible that the true rotation curve of NGC 891 is not flat, but slightly
falling down at large radii.\\
The results of this spectral cube modeling are listed below:   
\begin{eqnarray*}
     	\Theta_w(NGC \ 891)  & \approx &  90^{\circ}\\
	H_0(NGC \ 891) & \approx &  0.1 \ (kpc)\\
	\zeta_1(NGC \ 891) & \approx & 0.02\\
	R_f(NGC \ 891) & \approx & 12 \ (kpc)\\
	\zeta_2(NGC \ 891) & \approx & 0.02\\
	\alpha_{lin}(NGC \ 891) & \approx & 0.6-0.8\\
	R_w(NGC \ 891) & = & 12 \ (kpc)
\end{eqnarray*}
For comparison, we collected the HI parameters  
for the Milky Way, most of them coming from Merrifield (1992):
\begin{eqnarray*}
     	\Theta_w(MW)  & \approx &  80^{\circ}\\
	H_0(MW) & = &  0.1 \ (kpc)\\
	\zeta_1(MW) & \approx & 0.04\\
	R_f(MW) & \approx & 16 \ (kpc)\\
	\zeta_2(MW) & \approx & 0.08\\
	\alpha_{lin}(MW) & \approx & 0.3\\
	R_w(MW) & = & 12 \ (kpc)
\end{eqnarray*}
We can note significant differences between the two galaxies.
Although they both have the same 
constant HI layer thickness and both have a warp beginning at the same
radius (12 kpc). The warp of NGC 891 is however much higher in amplitude 
(reaching 6 to 8 kpc height at R=22 kpc, as the Milky Way's warp raises up
to 4 kpc height at R=25 kpc). Also the flaring of NGC 891 is less by a factor 2 
in comparison with the Milky Way. We have plotted (Fig~\ref{zirc90}) a
perspective view of our best fit model of NGC 891.    
\section{The flattening of an ellipsoidal DMH using the
outer flaring of the HI layer}
The fit of the rotation curve was made with an ellipsoidal dark matter 
density model. This axisymmetric dark matter density model has the advantage to
generate an axisymmetric potential that simplifies substantially the set of 
hydrodynamical and gravitationnal equations formed by the hydrostatic equilibrium 
of the HI gas and the Poisson equation. 
\dessin{8.5}{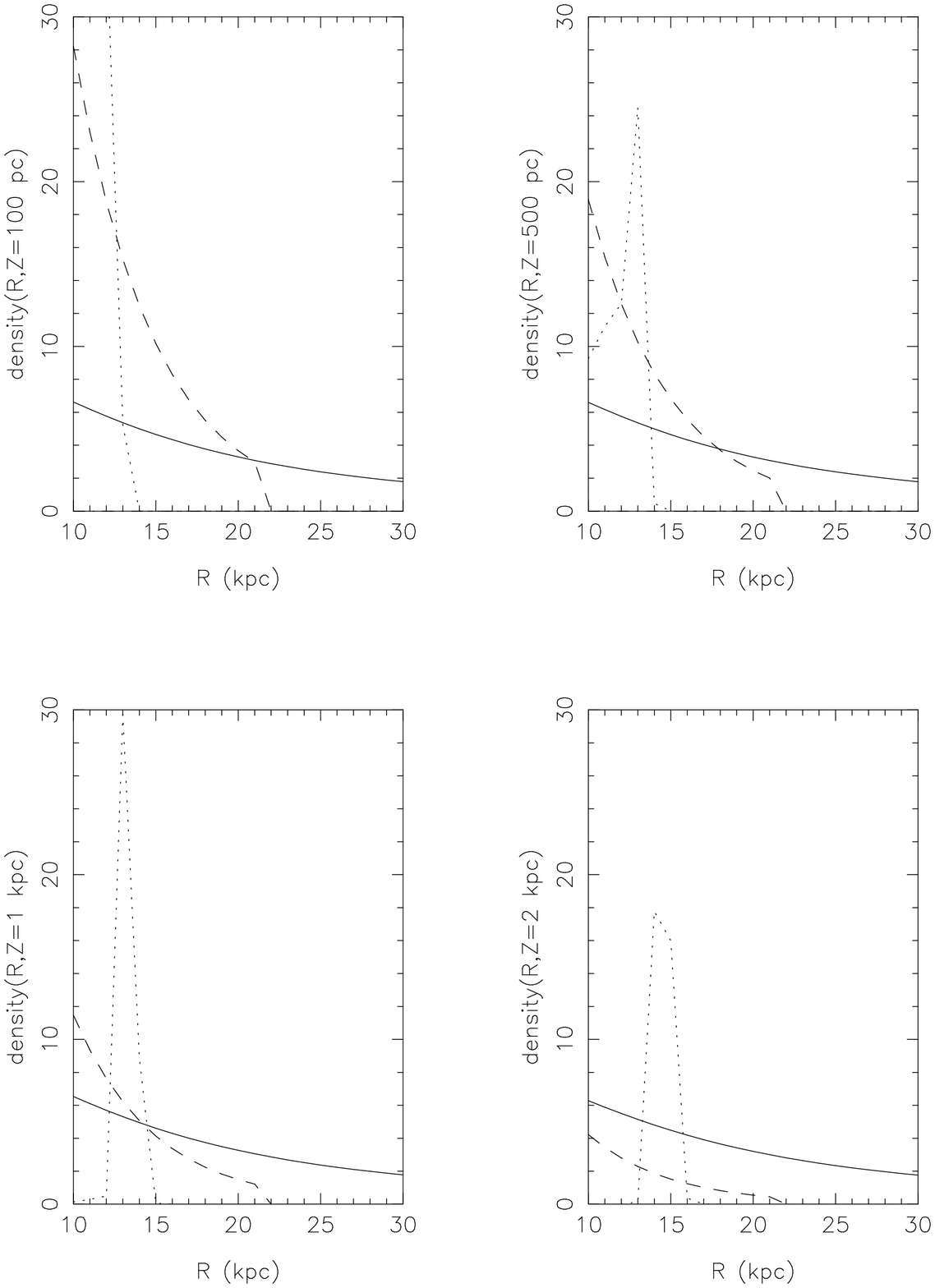}{We show here the densities of each component
of the galaxy, for different heights, and for radii corresponding to
the outer parts. The parameters of these components 
are found from the rotation curve fit.  
The dotted curve points out the HI density 
vertically distributed ($H_0=0.1 \ kpc$, $\zeta=0.02$) as a gaussian 
following the warp wave ($\alpha_{lin}(R_w=12kpc)=0.8$). 
The disk and DMH densities are 
dashed and full curves respectively. The DMH has fixed 
parameters: $R_c=14$ kpc and $q=0.5$. 
The bulge density for $R \ge 10$ kpc 
is totally negligible (and do not even appear in the graphs). 
The dark matter density term dominates after 20 kpc. 
Densities are in units of $0.001 \ M_{\odot}/pc^{3}$.}
{ro}
\dessin{8.5}{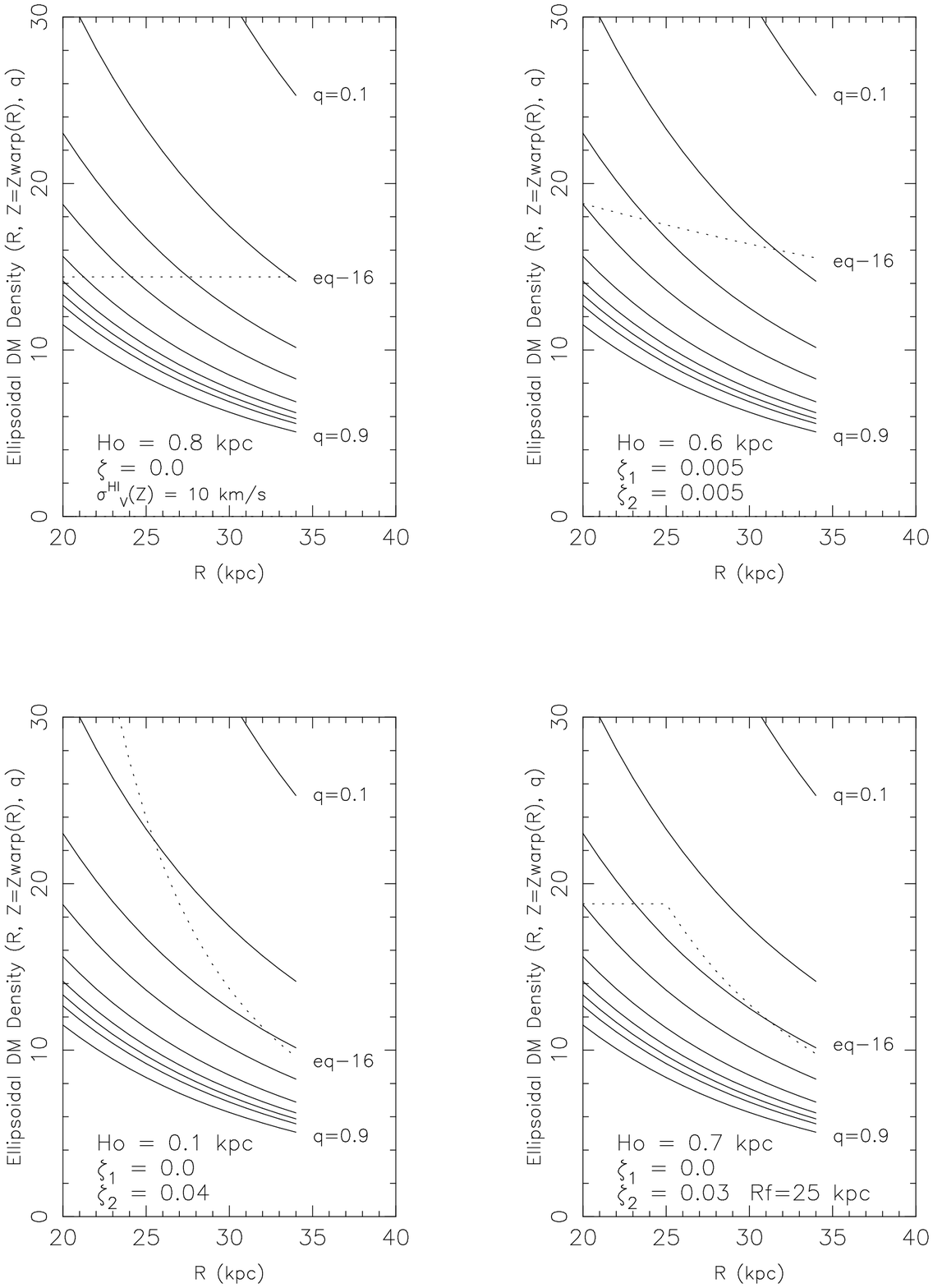}{We presents here plots of the right term of the 
equation 16 for different HI parameters (dotted curves).
The full lines are the left term of eq. 16, DMH densities 
$\rho_{DMH}$, corresponding to 
values of q from 0.1 to 0.9. We see that the HI term has very different
behaviours when we vary the $H_0$ and $\zeta$ parameters, so that the
equation can determine "q" unambiguously.
Densities are in units of $0.0002 \ M_{\odot}/pc^{3}$.}
{qdm0}
\dessin{7.5}{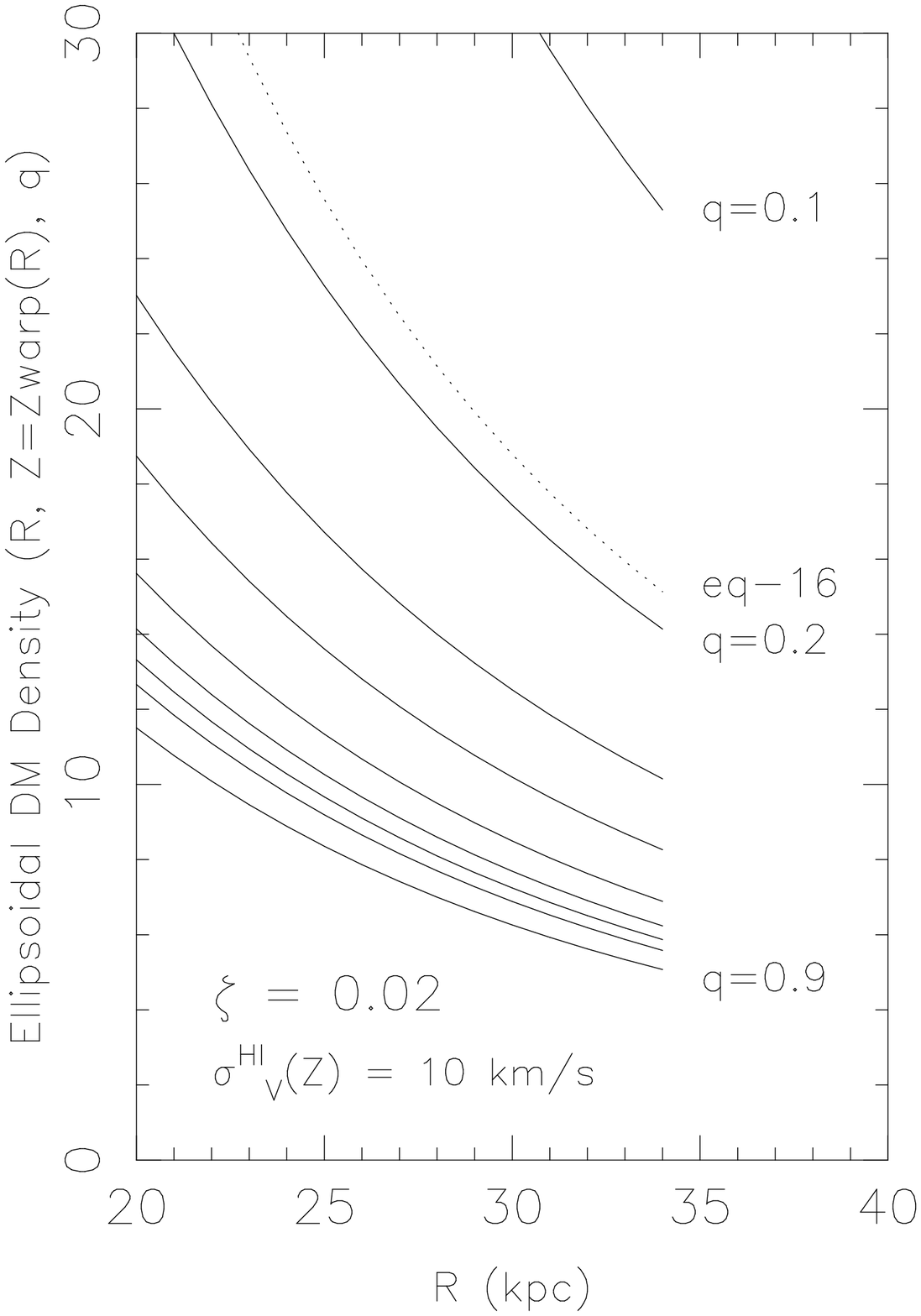}{We search here the solution of the 
equation (16) for the flaring value of NGC 891:
$\zeta_1=\zeta_2=\zeta=0.02$. We took here $\sigma_{HI}$
isotropic. 
The value $q=0.2$ is a good solution in average, the curve
being nicely "parallel" to the DMH density curves. 
Densities are in units of $0.0002 \ M_{\odot}/pc^{3}$.}
{qdm1}
\dessin{7.5}{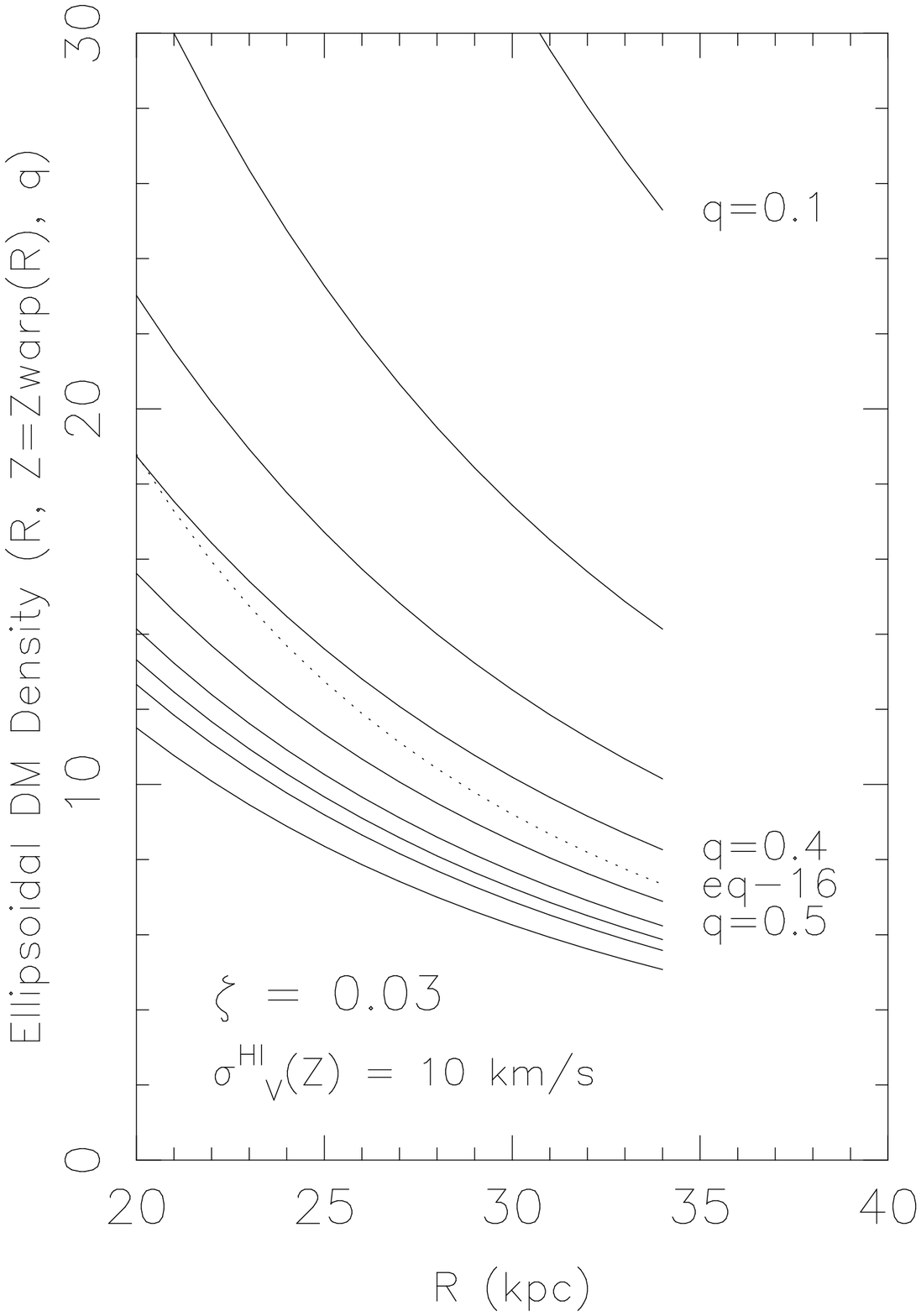}{Solution of the 
equation (16) for a greater value the flaring $\zeta = 0.03$, 
which could have been found without substracting the warp. 
We observe that this value modifies substantially the
flattening of the dark matter halo, pointing to an average
value of $q \approx 0.4-0.5$. The method is thus very sensitive
and the allowed range for $\zeta$ is between 0.015 and 0.035. 
Densities are in units of $0.0002 \ M_{\odot}/pc^{3}$.}
{qdm2}
\dessin{7.5}{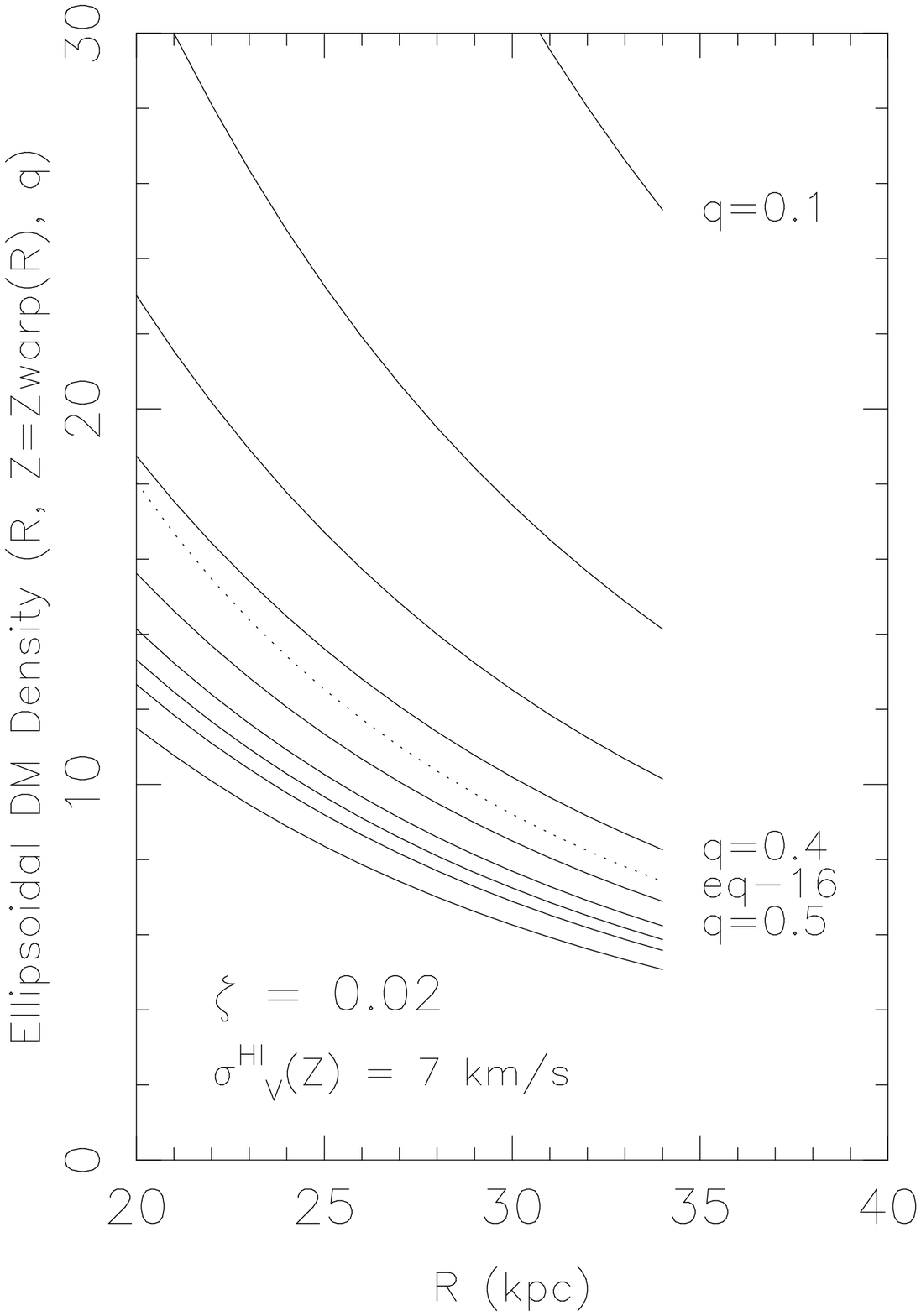}{Solution of the 
equation (16) for an anisotropic model of the HI velocity
dispersion: $\sigma_{HI,Z}=7$ km/s and 
$\sigma_{HI,R}=10$ km/s. We observe that 
it rounders substantially the DMH, pointing to an 
average flattening of $q \approx 0.4-0.5$. 
Densities are in units of $0.0002 \ M_{\odot}/pc^{3}$.}
{qdm3}
\\
Following Olling (1995),
the projection on the Z axis of the Jeans Equation becomes the
equation of Hydrostatic Equilibrium:
\begin{eqnarray}
\frac{ d(\rho(Z) \sigma_Z^{2}(Z))}{d Z} & = &
                    \rho(Z) F_Z(Z) - \frac{1}{R} \frac{\partial(
                     R \rho \sigma_{RZ}^{2})}{\partial R}
\end{eqnarray}
where the term in $\Theta$ has been eliminated due to the azimuthal
symmetry of the gaseous distribution ($m=0$ is overwhelming).
For $Z/R$ small ($Z \ge 0$) the tilt of the gaseous 
velocity dispersion ellipsoid $\sigma_{RZ}$ is approached by (BT87): 
\begin{eqnarray*}
\sigma_{RZ}^{2} & \approx & (\sigma_{R}^{2} - \sigma_{Z}^{2}) 
                          \frac{Z}{R}
\end{eqnarray*}
that we replace in the equation. 
Although $\sigma_Z$ cannot be measured at large R, observations of 
face-on external galaxies show little or no variation of the Z-dispersion
with R (e.g. Dickey et al, 1990).
Assuming the vertical and radial 
velocity dispersion of the HI gas independent of R:
\begin{eqnarray}
\sigma_{Z}^{2} \frac{d \ln \rho_{HI}}{d Z}  + 
(\sigma_{R}^{2} - \sigma_{Z}^{2}) \frac{Z}{R} 
\frac{d \ln \rho_{HI}}{d R} & = &
 -\frac{\partial \Phi_{galaxy}}{\partial Z}
\end{eqnarray}
where 
\begin{eqnarray*}
\Phi_{galaxy} & = &\Phi_{bulge} +\Phi_{\star} + 
                   \Phi_{HI} + \Phi_{DMH}
\end{eqnarray*}
Parallel to the hydrostatic equation (10), let us write the Poisson
equation for $R \ge 8$ kpc (flat region of the rotation curve).
Deriving with respect to R the equation of the flatness of the
rotation curve yields the condition:
\begin{eqnarray}
\frac{\partial^{2} \Phi_{galaxy}}{\partial R^{2}} + \frac{1}{R}
\frac{\partial \Phi_{galaxy}}{\partial R} & = & 0 
\end{eqnarray}
We insert (11) in the Poisson equation written in cylindrical
coordinates, which simplifies into (assuming the 
axisymmetry of the total galaxy potential):
\begin{eqnarray}
4 \pi G \rho_{galaxy} & = & \frac{\partial^{2} \Phi_{galaxy}}
                           {\partial Z^{2}}
\end{eqnarray}
We show in Fig~\ref{ro} the different densities in the outer parts, 
in order to simplify the term 
\begin{eqnarray*}
\rho_{galaxy} & = & \rho_{\star}+\rho_{bulge}+\rho_{HI}+\rho_{DMH}
\end{eqnarray*}
We see that the bulge density is clearly negligible, while after 
$R \ge 20$ kpc, the dark matter dominates largely the HI term.
We note that the dark matter density considered for the graphic 
corresponds to $q=0.5$; it is comparable for $q=0.8$, but
twice higher for $q=0.2$, so that the domination
of the dark matter in the outer parts is quite certain.
We may thus write 
\begin{eqnarray}
\rho_{galaxy}(R \ge 20 \ kpc) & \approx & \rho_{DMH}
\end{eqnarray}
Combining finally the vertical Jeans equation (10), 
the Poisson equation (12) and the density approximation (13) provides:
\begin{eqnarray}
\rho_{DMH} & = &    \frac{-\sigma_Z^{2}}{4 \pi G} \frac{\partial^{2} 
                                  \ln \rho_{HI}} {\partial Z^{2}}
				 + [\frac{-1}{4 \pi G}(\frac{(\sigma_R^{2} -
				 \sigma_Z^{2})}{R} \nonumber \\ 
                    &    &         (Z \frac{\partial^{2}
				 \ln \rho_{HI}}{\partial R \partial Z} 
                              + \frac{ \partial \ln \rho_{HI}}{\partial R})]
\end{eqnarray}
The term "$\ln \rho_{HI}(R,Z)$" is known since its
radial distribution has been derived in section 2.1 and its vertical distribution
$Z_0(R)$ has been modelled in previous sections:
\begin{eqnarray}
 \rho_{HI}(R,\Theta,Z)     & \approx & \rho_{0,HI}
                     \exp(-\frac{(R - R_g)^{2}}{2 \varrho_g^{2}}) \nonumber \\
                  &    & \exp(-\frac{(Z-Z_{warp}(R,\Theta))^{2}}{2 Z_0^{2}(R)}) 
\end{eqnarray}
The domain of validity of the equation (14) 
covers $R \sim 20-30$ kpc, where
the dark matter density dominates.
The term in bracketts in the right hand side of (14), coming 
from the tilt of the velocity
dispersion tensor, showed during computation to
be totally negligible (as expected from the observations, see 
Malhotra 1995). For $\sigma_R - \sigma_Z = 5$ km/s, 
this term represents no more than 1 \% of the first term.
So that inserting (15) in the simplified (14) provides finally:
\begin{eqnarray}
\rho_{DMH} (\rho_{0,DMH}(q),R \ge 20 \ kpc) 
              & = &  \frac{\sigma_Z^{2}}{4 \pi G Z_0^{2}(R)} 
\end{eqnarray}
We notice that the dark matter density term does not here depend on Z,
since the information on the flattening "q" is in fact included in
the central density $\rho_{0,DMH}(q)$, derived from the rotation curve
fit.
In the equation (16), the vertical velocity 
dispersion of the gas, $\sigma_Z$, 
carries much of the uncertainty. 
This quantity is frequently approached, in the literature, through
$\sigma_Z \approx \sigma_R$, but this is douftful. 
The interstellar
gas is under the form of small dense clumps, and the 
vertical component of its velocity dispersion
should be smaller than the radial component, in 
the same qualitative manner as the stars.\\ 
We draw in Fig~\ref{qdm0} the behaviour of 
equation (16) in its domain of validity, intending to show its
usefulness to derive q.
The figure Fig~\ref{qdm1} is the effective solution for our
problem (i.e. the value of $\zeta=0.02$ found in the previous spectral 
cube modeling section). We see that for a vertical velocity dispersion 
equal to the velocity dispersion in the plane, the average value of the
flattening is $q \approx 0.2$, pointing to a very flattened DMH.
We draw in  Fig~\ref{qdm2} the effects of changing the flaring up to a value
of $\zeta=0.03$: this rounders the DMH, with $q=0.4-0.5$. 
We draw in  Fig~\ref{qdm3} the effects of changing the vertical 
velocity dispersion of the gas to a lower value of 
$\sigma_{HI,Z} = 0.7 \ \sigma_{HI,R}$.
This makes also the DMH rounder at $q=0.4-0.5$.\\  
The method shows up tp be conclusive, reflecting small
variations of the $\zeta$-flaring coefficient and the vertical 
velocity dispersion.
We also note that there should exist a lower limit for the 
HI vertical velocity dispersion ($q \le 1.0$):
we find for NGC 891 that $\sigma_{HI,Z} \ge 5.5$ km/s.
A summary of the values of q as a function of flaring parameter
and HI z-velocity dispersion is listed in Table 2.
\begin{table}
\begin{flushleft}
\caption{Flattening of the DMH of NGC 891 ($R_c=14$ kpc)}
\begin{tabular}{l c l }
\hline
flaring   &   vertical dispersion   &  flattening\\
$\zeta$   & $\sigma_{HI,Z}$(km/s)   & q\\
\hline
0.02  & 10 & 0.2\\
0.03  & 10 & 0.4-0.5\\
0.02  & 7 & 0.4-0.5\\
\hline
\end{tabular}
\end{flushleft}
\end{table}
\section{Conclusion}
It was unfortunate that NGC 891 was found to have a warp
 which is maximum just along the line of sight, so that
the modeling of its HI flaring could not be as accurate as we could
hope.
Despite this difficulty, we have shown that it is possible to determine
 the HI plane thickness and the flaring  
for this quasi edge-on spiral galaxy using sequences of synthetic 
data cubes.
The warp of high amplitude for NGC 891 is also a nice explanation
for the observed strong variation with Z of the rotational velocities, a
behaviour detected previously by several authors.
We finally stress that the method seems efficient enough to constrain the
DMH flattening of every inclined galaxy 
for which we can determine the following quantities:\\
$\sigma_{HI,Z}$ [from a moment 2 map or face-on galaxies]\\
$R_c$, $\rho_{0,DMH}(q)$ [from fits of the rotation curve]\\
$H_0$, $\zeta$ [from a 3D spectral cube modeling]\\
Noteworthingly, the halo flattening found in this work is similar to the
result of R. Olling on NGC 4244, both suggesting very flattened halos. 
It thus may become now interesting to test
the dark matter model predicted by Pfenniger \& Combes, 
a warped disk of $H_2$ cold molecular gas.    
We shall present in a forthcoming paper the results of our method applied
to other galaxies with interferometric HI data cubes.
\begin{acknowledgements}
J.-F. Becquaert is grateful to Rob Olling, 
 Konrad Kuijken, Frederic Masset, Linda Sparke
and Roelof Bottema for useful and miscellaneous discussions 
about warps and halos. We specially thank Daniel Pfenniger for
his nice comments and  M. Rupen for the use of his
data of NGC 891 in digital form.
\end{acknowledgements}
\section{Appendix A: Non-cylindrical motions in galaxies}
Let us consider a very flattened
Kuzmin potential to represent the whole galaxy,
\begin{eqnarray*}
\Phi(R,Z) & = & \frac{G M}{\sqrt{R^{2}+(a_0 + |Z|)^{2}}}
\end{eqnarray*}
We expand $v_c^{2} = R \frac{\partial \Phi}{\partial R}$ assuming
$a_0 \ll Z \ll R$, obtaining:
\begin{eqnarray*}
v_c(z) & = & v_0 (1 - \frac{3}{4}(\frac{Z}{R})^{2})
\end{eqnarray*}
Applying this formula for NGC 891, at $R=10$ kpc and $Z=3$ kpc 
gives for $v_0=225$ km/s  
a loss of velocity of about 15 km/s.
A vertical gradient of velocity is thus generally expected in
galaxies.
We include explicitly this gradient in the model
in multiplying the disk velocity by a factor $f_V(Z)$, which is chosen
in practice to be $(1 - \frac{3}{4}(\frac{Z}{R})^{2})$.
\section{Appendix B: Derivation of the spiral arm velocities}
We entered in the modeling of the spiral arms their perturbed 
velocities associated with the mode $m$ (practically, $m=2$).
Using Euler's equation, BT87 estimated that $V_R$ and $V_{\Theta}$ are
of the same order and thus the continuity equation could be simplified,
(pages 356-359) resulting in the formula of the perturbed radial 
velocity associated with the mode $m$: 
(under adaptations for the present context)
\begin{eqnarray*}
V_R(m) & = & - \frac{m \Omega - \omega}{k_m} \ \frac{A_m}{A_0} \ 
             [\frac{1}{R} \cos (k_m R + m(\Theta -\Theta_m))]
\end{eqnarray*}
The perturbed tangential velocity associated with the mode $m$ 
is obstained from formula (6-36)
of BT87, yielding:
\begin{eqnarray*}
V_{\Theta}(m) & = & \frac{- 2 B}{k_m} \ \frac{A_m}{A_0} \
                    [\frac{1}{R} \sin (k_m R + m(\Theta -\Theta_m))]
\end{eqnarray*}
 "B" being Oort's second "constant".
It is noteworthy that $V_R$ changes sign at corotation, while
$V_{\Theta}$ does not. 
The kinematical parameters a-priori unknown in the formula above
are derivable from a multi-component fit of the rotation
curve (Fig~\ref {vfit1}): the value of $\Omega = \frac{V_{rot}}{R}$,
 $\kappa = \sqrt{R \frac{d\Omega^{2}}{dR} + 4\Omega^{2}}$
 and $B=B(R)=-\frac{\kappa^{2}}{4 \Omega}$.
The main modification in the spectral cube construction due to the
consideration of these perturbed velocities, is to change the
formula of the line-of-sight velocity, which is now:
\begin{eqnarray*}        
V_{los}(m=0) & = & V_{sys} + V_{rot} \sin I \cos \Theta f_V(Z)\\
V_{los}(m \neq 0) & = & V_{sys} + [(V_{rot}+V_{\Theta}(m)) 
                        \sin I \cos \Theta \\
                  &   & + V_R(m) \sin I \sin \Theta] f_V(Z)
\end{eqnarray*} 
\section{Appendix C: potentials for generalized dark matter densities}
Simulations of spherical infall in an Einstein-de Sitter
universe onto seeds perturbations and onto density peaks in
power cosmologies result in objects with $\rho \propto R^{- \alpha}$,
$1.6 \le \alpha \le 2.25$ (Dubinski \& Carlberg, 1991). 
This is close enough to the isothermal profile 
(i.e. $R^{-2}$), which is often adopted for the sake of simplicity.
However, Lake \& Feinswog (1989) showed that profiles in $R^{-3}$ 
and even $R^{-4}$ were also possible to obtain a good fit of
observed rotation curves. Taking this result into account, we consider 
hereafter the generalized family of densities: 
\begin{eqnarray*}
\rho_p(q) & = & \frac{\rho_{0,DMH}}{(1+ \frac{R^{2}}{R_c^{2}} 
                 +\frac{Z^{2}}{R_c^{2} q^{2}})^{p}} \ (p \ge 0)
\end{eqnarray*}
This density model recovers,
for $p=1$, the formula of the isothermal
density profile used in this article and previously by other authors.
We give here the formula of the dark matter potential for $p \ne 1$, which
 can be derived using formula 2.6 of de Zeeuw \& Pfenniger (1988),
and the new variable ($x=\frac{R_c}{\sqrt{u+q^{2}R_c^{2}}}$), providing:
\begin{eqnarray*}
\Phi_{DMH}(R,Z) & = & - \frac{2 \pi}{1-p} G \rho_{0,DMH} q R_c^{2} \\ 
                   &   & \int_{0}^{\frac{1}{q}} 
                          \frac{ [1 + \frac{x^{2}}{R_c^{2}} 
                           (\frac{R^{2}}{\epsilon^{2}
                             x^{2} + 1} + Z^{2})] ^{1-p}}{\epsilon^{2}
                             x^{2} + 1} dx
\end{eqnarray*}
The rotation curve contribution is, from that moment, straightforward.
\end{document}